WILEY-VCH

# Quantifying Nonradiative Recombination and Resistive Losses in Perovskite Photovoltaics: A Modified Diode Model Approach

*Minshen Lin[1], Xuehui Xu[2], Hong Tian[2], Yang (Michael) Yang[2]\*, Wei E. I. Sha[3]\*, and Wenxing Zhong[1]\**

[1]College of Electrical Engineering, Zhejiang University, Hangzhou 310027, China.

[2]State Key Laboratory of Modern Optical Instrumentation, Institute for Advanced Photonics, College of Optical Science and Engineering, Zhejiang University, Hangzhou 310027, China

[3]State Key Laboratory of Modern Optical Instrumentation, College of Information Science and Electronic Engineering, Zhejiang University, Hangzhou 310027, China

\*Authors to whom correspondence should be addressed: yangyang15@zju.edu.cn, weisha@zju.edu.cn, and wxzhong@zju.edu.cn

## Abstract

Pinpointing the origin of inefficiency can expedite the process of optimizing the efficiency of perovskite photovoltaics. However, it is challenging to discern and quantify the different loss pathways in a complete perovskite photovoltaic device under operational conditions. To address this challenge, we propose a modified diode model that can quantify bulk/interface defect-assisted recombination and series/shunt resistive losses. By adopting drift-diffusion simulation as the benchmark, we explore the physical meanings of the modified diode model parameters and evaluate the performance of the model for simulation parameters spanning many orders of magnitude. Our evaluation shows that, in most practical cases, the proposed model can accurately quantify all the aforementioned losses, and in some special cases, it is possible to identify the predominant loss pathway. Moreover, we apply the modified diode model to our lab-produced devices (based on $Cs_{0.05}FA_{0.95}PbI_3$ perovskites), demonstrating its effectiveness in quantifying entangled losses in practice. Finally, we provide a set of guidelines for applying the modified diode model and interpreting the results. Source code available at https://github.com/WPT-Lab124/Modified-Diode-Model.

## 1. Introduction

With over ten years of development, perovskite photovoltaics have shown tremendous improvement in both efficiency[1] and stability[2]. The record efficiency of single-junction perovskite photovoltaics has reached 25.8%, rivaling that of crystalline silicon photovoltaics[3]. However, such efficiency is still below the thermodynamic efficiency limit of 31%, as predicted by the Shockley-Queisser formulation[4]. To improve the performance of perovskite photovoltaics, strategies involving modification of materials or even device structures are often adopted[5]. In this respect, during the exploration of a more performant device, it is not uncommon for the modified devices to fall short of the expected efficiency. This requires devising an effective strategy that can facilitate targeted efficiency improvement, which relies on the understanding and identification of each loss mechanism inside the device.

The losses in photovoltaics fall into three categories—optical loss, recombination loss, and resistive loss[6]. Recombination and resistive losses are also known as the electrical losses. Under 1-Sun irradiance, the performance of perovskite photovoltaics is predominantly limited by the defect-assisted recombination either in the perovskite bulk or at the perovskite/transport layer interfaces, while under higher irradiances or in devices with comparatively large areas, resistive loss can predominate[7]. These losses are coupled via carrier kinetics such that disentangling one from the others demands detailed modeling of the device behavior. In the literature, experimental techniques such as steady-state photoluminescence[7–9], time-resolved photoluminescence (TRPL)[7–10], impedance spectroscopy[11,12], and ideality factor measurement[13–18] have been used to analyze the non-radiative recombination losses. More recently, a light-intensity approach has been proposed to help in the explanation of electrical losses[5]. Though these methods are useful guides in particular cases, they cannot provide comprehensive information regarding all the electrical losses, or quantify each loss.

Generally, drift-diffusion (DD) simulations[19–21] and circuit model analysis[22–25] are used to analyze and quantify the losses in photovoltaics, with necessary parameters retrieved by experimental techniques. In DD simulations, a large number of parameters are required and accurate determination of some critical parameters such as interface recombination velocity is challenging; hence, such modeling is inconvenient and sometimes difficult to apply during the iterations of device optimization. On the other hand, circuit models can analyze the performance with measured current density-voltage ($J$-$V$) curves, via curve fitting, which is easy to implement. However, in classic circuit models such as single- and double-diode models[24,25], the parameters regarding dark current density and ideality factor have ambiguous physical meanings such that they cannot be directly associated with the recombination processes in perovskite photovoltaics. Besides, in the recently developed circuit models for perovskite photovoltaics[22,23], the focus is on the explanation of internal charge distribution and ideality factor, instead of loss quantification.

Aiming at both accurate loss modeling and clear physical meaning, in this study, we introduce and analyze the modified diode (MD) model, applied in such a way that it can discern and quantify the following losses: 1) Shockley-Read-Hall (SRH) recombination loss occurred in the perovskite bulk; 2) SRH recombination loss occurred at the perovskite/transport layer interfaces; 3) series resistance loss that usually results from the lateral conduction in the transparent electrode and the geometry of the device[5,7,26]; 4) shunt resistance loss due to the defects and voids-induced current leakage[6]. By adopting DD simulation as the benchmark, first, we explore the physical meanings of each parameter in the MD model. Then, we comprehensively evaluate the accuracy of the MD model for device parameters (in DD simulation) spanning many orders of magnitude that cover most of the practical cases. Such evaluation also includes high irradiance applications, considering that perovskite photovoltaics are potential candidates for concentrator[27,28] and optical wireless power transfer[29,30] technologies. Moreover, we analyze how mobile ions in the perovskite layer can affect the accuracy of the model in describing recombination currents, and then apply the MD model to our lab-produced devices, demonstrating its capability in quantifying entangled losses in practice. Finally, we provide a set of guidelines for applying the MD model and interpreting the results. We make the source code (implemented in MATLAB) open-source online[31], which can hopefully facilitate the design of efficiency-targeted optimization strategies for researchers in this field.

## 2. Modeling and Evaluation

### 2.1. Theory of Modified Diode Model

The MD model draws certain characteristics from both the detailed balance model and the circuit models for photovoltaics, providing a way to discern recombination and resistive losses in practical devices. In the detailed balance model, ideal contacts and infinitely high carrier mobilities are assumed for the limiting case so that the quasi-Fermi level splitting (QFLS) equals $qV$—the product of the elementary charge, $q$, and the applied voltage, $V$. In this way, the radiative recombination current can be described in terms of the QFLS (as a function of $V$), connecting the external applied voltage with the internal carrier kinetics. To utilize similar relations and to incorporate resistances as with circuit models, in the MD model, we attribute the resistive losses produced by lateral conduction in the transparent electrode and geometry effects[5,7,26] to a lumped series resistance, $R_s$, such that the QFLS approximates $q(V + JR_s)$, where $J$ is the output current density. With this generalization and to account for non-radiative recombination losses, the current density-voltage ($JV$) relation is given by

$$J = J_{ph} - J_r(V + JR_s) - J_{nr}(V + JR_s) - \frac{V + JR_s}{R_{sh}}, \tag{1}$$

where $J_{ph}$ is the photocurrent density, $J_r$ is the radiative recombination current density, $J_{nr}$ is the non-radiative recombination current density, and $R_{sh}$ is the shunt resistance, representing the defects and voids-induced current leakage[6]. The photocurrent density can be calculated by the absorption of the illumination spectrum; likewise, the radiative recombination current density is determined, according to the principle of detailed balance, by the absorption of the dark spectrum.

Of greater interest in practical devices is the non-radiative recombination, which is comprised of several components representing different recombination pathways. In perovskite photovoltaics, recombination losses are predominantly due to the Shockley-Read-Hall (SRH) recombination either in the perovskite bulk or at the perovskite/transport layer interfaces[7]. Second-order radiative recombination and third-order Auger recombination may predominate with higher carrier concentrations. However, as assessed by an extraction-dependent rate equation model, radiative and Auger recombination prevail only when the illumination irradiance is over several hundreds of Suns[28]. In this work, we analyze the performance of perovskite photovoltaics up to 50 Suns; in such cases, radiative and Auger recombination currents are smaller than SRH recombination currents by several orders of magnitude. Thereby, we only consider the currents due to SRH recombination in Eq. (1). In order to differentiate between bulk and interface SRH recombination, we introduce two recombination current densities of similar form into the MD model,

$$J_{SRH}^{bulk} = qL\gamma_{bulk}n_i\left[\exp\left(\frac{V + JR_s}{2k_BT/q}\right) - 1\right], \tag{2}$$

$$J_{SRH}^{if} = qU_{if}\left[\exp\left(\frac{V + JR_s}{n_{id}^{if}k_BT/q}\right) - 1\right], \tag{3}$$

where $\gamma_{bulk}$ is the bulk SRH recombination rate, $L$ is the active layer thickness, $n_i$ is the intrinsic carrier density of the active layer material, $U_{if}$ is the interface SRH recombination rate, and $n_{id}^{if}$ is the ideality factor for interface SRH recombination. The detailed derivation of these equations is given in Supplementary Note 1, along with the necessary assumptions. Later sections of this paper discuss the justification of these assumptions in more detail. Besides, one reason why we fix the value of the ideality factor for bulk SRH recombination is that if the ideality factors for both recombination pathways are allowed to vary, then interpreting bulk SRH recombination current as interface SRH recombination current or vice versa is of no difference in terms of curve fitting, which can lead to questionable uniqueness of the retrieved parameters.

## 2.2. Simulation Study

To explore the accuracy of the MD model, we use DD simulation results as the benchmarks, assessing how well Eq. (2) and Eq. (3) capture the recombination currents. The assessment procedure is described in the following:

1) First, the well-established DD simulator, SCAPS-1D[32], is used to produce the $JV$ curves—more precisely, the $JV$ data points.

2) Then we retrieve the unknown parameters in the MD model such that the reproduced $JV$ curves (data points) have minimum residuals with respect to the simulation results. This is essentially a curve-fitting problem. Since $R_s$ and $n_{id}^{if}$ are in the exponential terms, we leverage the Levenberg-Marquardt (LM) algorithm[33,34] to solve such non-linear least squares problem.

3) Finally, we use the retrieved parameters to solve the recombination currents, Eq. (2-3), for comparison with the DD simulation results.

To begin with, we consider the pin-type device consisting of indium tin oxide (ITO, 150 nm)/PTAA (8 nm)/perovskite (700 nm)/C60 (30 nm)/Ag (100 nm). The parameter values for DD simulations are mostly adopted from the literature[7,14], in which the bulk SRH recombination lifetime and interface SRH recombination velocity are extracted from photoluminescence experiments. Unless particularly specified, the parameter values in the rest of the article are consistent with those listed in Table S1. In addition, we include a series resistance of $1\ \Omega \cdot cm^2$ and a shunt resistance of $10^7\ \Omega \cdot cm^2$ to model the resistive losses. For the purpose of clearly interpreting the MD model, we introduce three regions—low-voltage region, exponential region, and consumer region—to describe the operation of photovoltaics specified by different voltages. The low-voltage region refers to the part in the $JV$ curve where the current density almost stays constant with increasing applied voltage, starting from the voltage corresponding to the photocurrent. The exponential region refers to the other part where the current density grows drastically with applied voltage[15]. We choose the maximum power point (MPP) voltage to distinguish between these two regions for 1-Sun irradiance, and for higher irradiances, we choose the voltage where the exponential tail starts. Besides, we refer to the region where the applied voltage is greater than the open-circuit voltage as the consumer region since in this region the device consumes energy.

At the first attempt, we allow all the five unknown parameters in the MD model ($\gamma_{bulk}, U_{if}, n_{id}^{if}, R_s, R_{sh}$) to vary within a broad range (from 0 to $10^{12}$) in the course of curve fitting. Figure 1 (a) shows the DD simulated and MD model fitted $JV$ curves. These two $JV$ curves exhibit negligible fitting errors, which, however, does not guarantee that the MD model

## WILEY-VCH

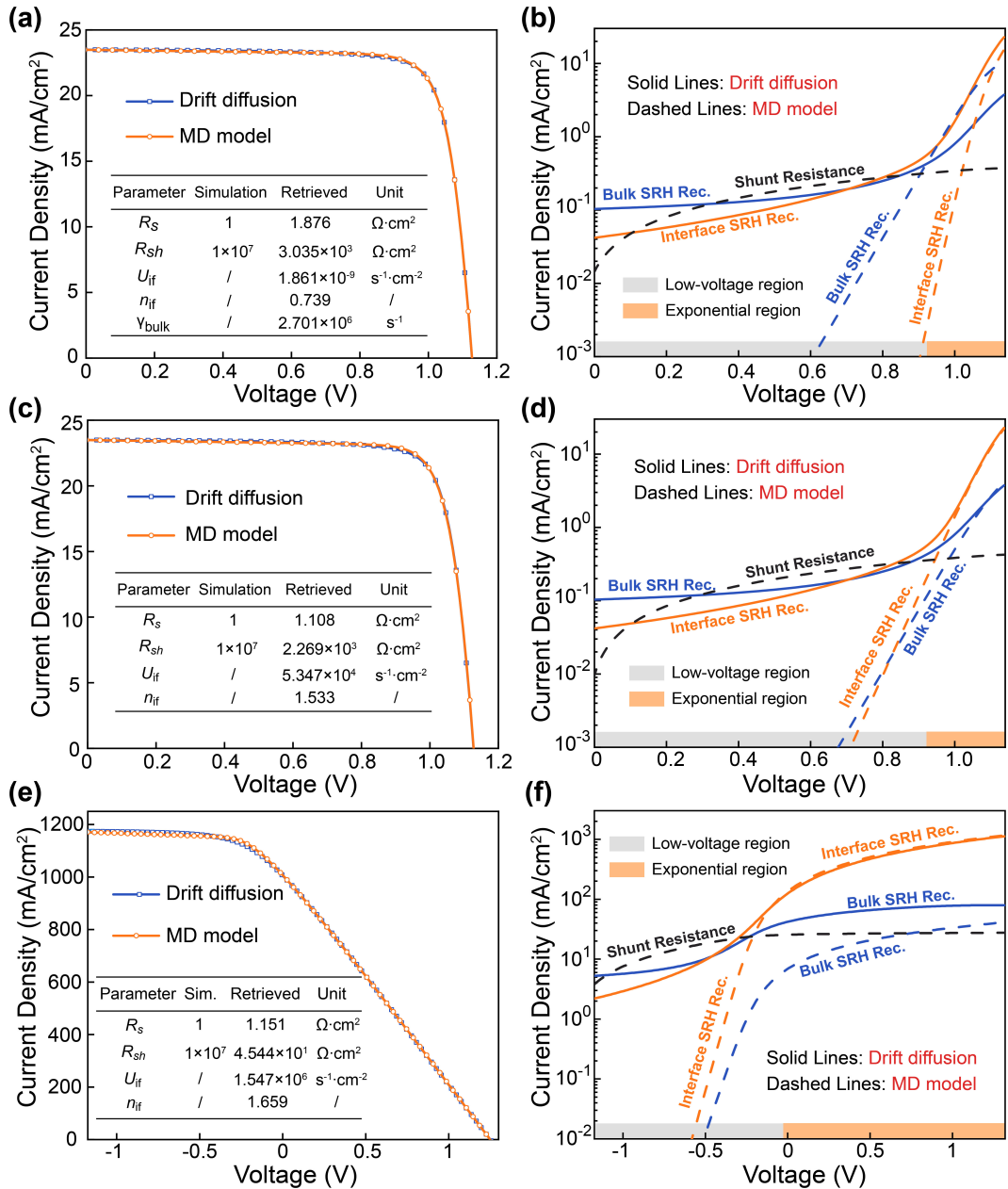

**Figure 1.** Current density-voltage relations of perovskite photovoltaics. **(a-b)** Curve fitting with the MD model, allowing all the five parameters listed in the inset table to vary. **(c-f)** Curve fitting with the MD model, allowing the four parameters listed in the inset table to vary. Drift diffusion (DD) simulation and modified diode (MD) model fitting results are given in **(a)**, **(c)**, and **(e)**. The current densities due to different loss pathways calculated by the MD model using retrieved parameters are compared with the corresponding simulation results in **(b)**, **(d)**, and **(f)**. The irradiance of illumination (AM 1.5G spectrum) is 1 Sun in **(a-d)** and 50 Suns in **(e-f)**.

captures all the information. In Fig. 1(b), the simulated and fitted recombination currents are depicted. In the exponential region, it is evident that the fitted recombination currents do not correctly describe the corresponding simulated recombination currents; instead, the fitted bulk SRH recombination current accounts for the predominant simulated interface SRH recombination current, which can lead to false loss analysis. To tackle this problem, we leverage the bulk SRH recombination lifetime. The bulk SRH recombination rate, $\gamma_{bulk}$, derived from the SRH recombination formulation, equals to $1/2\tau_{bulk}$ in case of mid-gap traps and $n = p$, where $\tau_{bulk}$ is the SRH recombination lifetime. This lifetime is an input parameter in the DD simulations, and in practical devices, can be measured by TRPL experiments[7,10]. Given a known $\tau_{bulk}$, we can fix the value of $\gamma_{bulk}$ and only retrieve the values of the remaining four parameters ($U_{if}, n_{id}^{if}, R_s, R_{sh}$) by curve fitting. Using this strategy, we implement the procedure again. Fig. 1(c) shows the newly fitted $JV$ curve with retrieved parameter values listed in the inset table. Noticeably, the fitted recombination currents, depicted in Fig. 1(d), accurately follow the respective simulation results in the exponential region. This fitting result suggests that the MD model is capable of describing and discerning different SRH recombination mechanisms, so long as the bulk SRH recombination lifetime is given. Similarly, we implement the assessment procedure to the same device but increase the illumination irradiance to 50 Suns. As shown in Fig. 1(e-f), the MD model well captures the predominant interface SRH recombination current. Though the fitted bulk SRH recombination current is less than the simulation result, this introduces negligible error in terms of loss analysis, because the bulk SRH recombination only takes up a small portion of the overall loss.

So far, we have not discussed the resistive components, which are also of great importance in many devices. As shown in the inset table in Fig. 1, the retrieved values of $R_s$ are slightly greater than the given value in the simulation. Potentially, this error originates in the enhanced recombination loss due to non-ideal transportation, which is not associated with the external series resistance that we additionally include. On the other hand, the retrieved values of $R_{sh}$ seem to deviate far from the given value in the simulation. Yet from Fig. 1(d) and Fig. 1(f), we can see that in the MD model, the current density that flows through $R_{sh}$ serves as a compensation for the SRH recombination current in the low-voltage region, since in this region, the recombination current does not exponentially grow with applied voltage. The increase of recombination current with respect to applied voltage can be understood in terms of the accumulated carrier concentrations in the absorption layer: 1) Under illumination, the photo-generated carriers are not completely extracted due to finite carrier mobilities, part of which accumulate in the absorption layer contributing to recombination. 2) At relatively low applied voltages, the carrier concentration in the absorption layer is the sum of the predominant accumulated photo-generated carrier concentration and the voltage-injected carrier concentration, which have a slower-than-exponential (almost linear) increase with respect to the applied voltage. Since this increase approximates a linear process, in the MD model, it is naturally attributed to the shunt resistance by curve fitting. 3) With increasing applied voltage,

WILEY-VCH

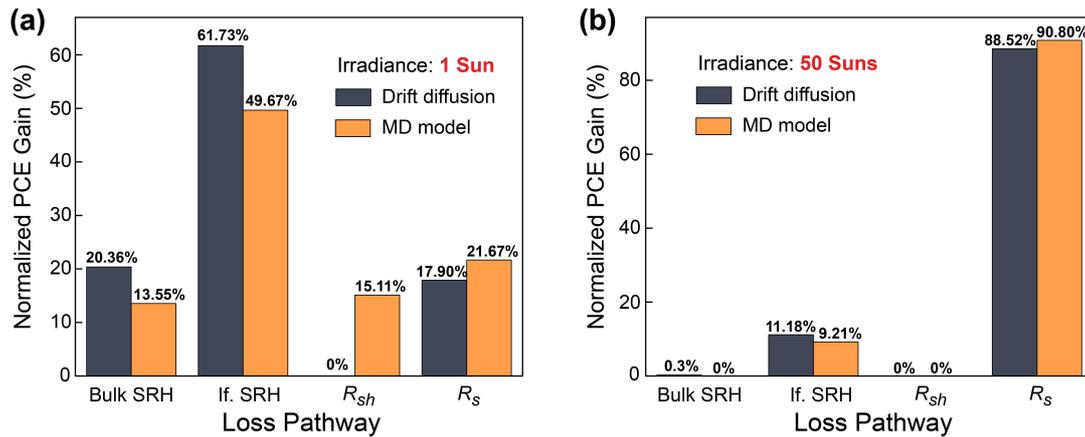

**Figure 2.** Loss analysis using both drift-diffusion simulation (DD) and modified diode (MD) model for **(a)** 1 Sun irradiance and **(b)** 50 Suns irradiance. The impact of each loss pathway is illustrated in terms of the normalized PCE gain, which is determined by the PCE gain excluding that loss.

the electrically injected carriers gradually form a background concentration that outweighs the accumulated photo-generated carriers such that the carrier concentration and the recombination currents follow the exponential increase with respect to voltage. These behaviors can also be observed in Fig. S1, where carrier distribution and QFLS at different voltages are demonstrated. We can therefore confirm that $R_{sh}$ in the MD model takes on an additional role in describing the recombination currents in the low-voltage region, which increases non-exponentially with applied voltage.

Using the retrieved parameters, we can quantitatively evaluate the electrical losses due to different recombination pathways and resistances. However, the impact of each loss pathway is not independent of the others. For example, compared to smaller series resistance, a larger one not only produces a higher resistive loss, but can also build up larger QFLS inside the device, which inevitably leads to higher recombination losses. Thus, simply calculating each loss in terms of power consumption (e.g., at maximum power point) cannot provide useful information. To separate the impact of each loss mechanism, we evaluate the power conversion efficiency (PCE) gains by excluding one specific loss pathway at a time. Figure 2 depicts the results of loss analysis by both DD and MD models, where we individually exclude bulk SRH recombination, interface SRH recombination, shunt resistance, and series resistance to calculate the corresponding PCE gains. Moreover, to clearly present the results that indicate the relative importance of each loss pathway, the PCE gains are normalized to unity. The details of such calculation are provided in Supplementary Note 3. In Fig. 2(a), under 1 Sun irradiance, the normalized PCE gain predicted by the MD model shows some noticeable errors with respect to the DD simulation results. In fact, these errors are quite small before normalization. For example, the absolute PCE gain by excluding the bulk SRH recombination is 0.55% in the DD model and 0.40% in the MD model; the absolute PCE gain by excluding the interface SRH recombination is 1.66% in the DD model and 1.47% in the MD model. Therefore, we can still use the MD model to predict the PCE gain with confidence, and the major reason to use normalized data is to show their relative importance. Besides, the origin of these errors is traceable. As listed in Fig. 1, the MD model tends to retrieve a slightly larger value of $R_s$, which can lead to a small overestimation of PCE gain by excluding $R_s$. Likewise, having a slightly larger $R_s$ during calculation can in turn underestimate the PCE gains by excluding other recombination pathways. We also notice that the PCE gain due to excluding $R_{sh}$ is only visible in the MD model. As analyzed previously, the shunt resistance describes the recombination currents in the low-voltage region, which is different from the external shunt resistance in the DD simulations that models the current leakage; thereby, excluding $R_{sh}$ in the MD model in effect excludes that certain amount of recombination, leading to the efficiency gain. Despite such an error, we should still keep track of the shunt resistance because, in some devices where current leakage is severe, the PCE gain calculated by the MD model can produce consistent results. An example is shown in Fig. S2 where a small $R_{sh}$ of $100\ \Omega \cdot cm^2$ is introduced into the simulation for modeling current leakage; in this case, the shunt current described by the MD model corresponds exactly to the current leakage. Figure 2(b) depicts the normalized PCE gain under 50 Suns irradiance. The results show very small errors compared to the 1 Sun case, meaning that the MD model accurately determines the relative importance of each loss pathway. Importantly, for both irradiances, the crucial information of the efficiency limiting factors, in a relative or absolute manner, is successfully predicted by the MD model. Therefore, the MD model is capable of providing valuable design guidance towards higher efficiency devices, for both 1 Sun and high irradiance applications.

There is one more parameter in the MD model, $n_{id}^{if}$, which requires further scrutiny. Normally, the ideality factor of a photovoltaic device is determined by the Suns-$V_{oc}$ method rather than curve fitting[14,15,17]. These two methods can give distinct values of ideality factors for the same device. Here, we explore the physics behind the ideality factor retrieved by curve fitting, which turns out to be an important reason why the MD model performs as expected. In the MD model and other circuit models, the ideality factor is given as a voltage-independent parameter. The curve-fitting process picks up the value for the ideality factor such that the recombination current of the exponential form fits the simulated recombination current. With this notion, we can calculate the differential ideality factor for each simulated recombination current by perceiving them as Eq. (2-3) at every data point. Namely, for the entire $JV$ curve, the recombination current is assumed to have the form

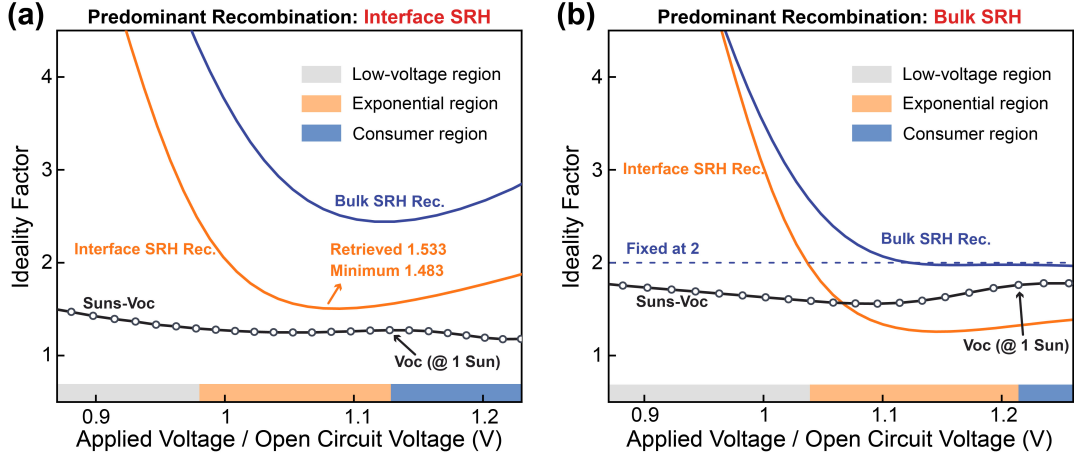

**Figure 3.** Ideality factor as a function of voltage calculated by both Suns-$V_{oc}$ method and Eq. (5). **(a)** The analyzed device has parameters listed in Table S1, inside of which the interface SRH recombination predominates. **(b)** The same device but with reduced interface recombination velocity ($S_{if} = 1$ cm/s) is analyzed, inside of which the bulk SRH recombination predominates.

$$J_{rec} = J_o \exp\left[\frac{q(V + JR_s)}{n_{id} k_B T}\right],\tag{4}$$

where $J_{rec}$ denotes the recombination current density of any type, $J_o$ is the dark current density, and $n_{id}$ is the ideality factor. Though such perception is not accurate in the low voltage region where the carrier concentration is not predominated by voltage injection, the fitting results in Fig. 1 indicate that the exponential region is of most relevance to the recombination currents in the MD model, whereas $R_{sh}$ describes the low-voltage region recombination. This also suggests that so long as the ideality factor in the exponential region is not a strongly voltage-dependent parameter, we can retrieve a constant value of the ideality factor that reflects the average information of the exponential growth. Based on the previous studies[14,15,17], we additionally take series resistance into account to calculate the phenomenological differential ideality factor,

$$n_{id} = \frac{q}{k_B T}\left(\frac{d \ln J_{rec}}{dV}\right)^{-1}\left(1 - \frac{dJ_{rec}}{dV} R_s\right).\tag{5}$$

The detailed derivation is given in Supplementary Note 4. With Eq. (5), we can calculate $n_{id}$ as a function of applied voltage for both interface and bulk SRH recombination currents, for the same device under 1 Sun illumination. Figure 3 depicts the $n_{id}$ for each simulated SRH recombination current (as those shown in Fig. 1(d)), in addition to the ideality factor determined by the Suns-$V_{oc}$ method. In agreement with the above analysis, the retrieved value of $n_{id}^{if}$, 1.533, which is slightly greater than the minimum differential ideality factor, lies in the exponential region of the $JV$ curve, capturing the average information of the exponentially increasing recombination currents. Further, we reduce the interface recombination velocity to 1 cm/s, in order to simulate a device where bulk SRH recombination predominates. As shown in Fig. 3(b), the corresponding ideality factor for the bulk SRH recombination current turns out to be very close to 2 in the exponential region, justifying our approach to using a fixed ideality factor for bulk SRH recombination current. There are two reasons that underline the emergence of this constant ideality factor: 1) In the exponential region, the carriers in the bulk of the perovskite are predominated by electrical injection, whose product follows an exponential increase with QFLS according to the Boltzmann statistics, i.e., $np = n_i^2 \exp(\text{QFLS}/k_B T)$ . 2) With more carriers injected by increasing voltage, the electrons and holes tend to establish charge neutrality in the perovskite bulk, i.e., $n = p$. This can be explained by the decrease in Debye length, $L_D = \sqrt{\varepsilon k_B T / q^2 n}$, which reflects the ability of carriers to screen out net charge and establish charge neutrality[35]. The higher the concentration of carriers, the shorter the Debye length and the better the balance in electron and hole concentrations. Such balance, combined with the carrier statistics, allows us to reduce the bulk SRH recombination current to a simpler form where the ideality factor is 2 (Supplementary Note 2). Besides, in both Fig. 3(a) and Fig. 3(b), we notice that the differential ideality factor of the predominant recombination current is greater than that obtained by the Suns-$V_{oc}$ method, which is attributed to finite carrier mobilities and resistive effects in the literature[15,17]. Therefore, the above analysis shows that the retrieved ideality factor by curve fitting has a clear physical meaning related to how recombination currents grow in an average sense, and the discrepancy between the values in our work and those in the literature[14] is reasonable since those values are derived from the Suns-$V_{oc}$ method.

### 2.3. Accuracy Evaluation

In general cases, perovskite photovoltaics can have various parameters, different material compositions, and even distinct device structures such as nip type. In order to evaluate the performance of the MD model applied to general cases, we introduce a cost function that adds up a particular form of residuals,

$$Cost = \sum_{i=1}^{N} \frac{|J_{DD} - J_{MDB}| + |J_{DD}^{rec} - J_{MDB}^{rec}|}{N \times J_{ph}}, \tag{6}$$

where $J_{DD}$ and $J_{MDB}$ are the output current densities, $J_{DD}^{rec}$ and $J_{MDB}^{rec}$ are the SRH recombination current densities, and $N$ is the number of data points. The subscripts, *DD* and *MD*, denote that the current density is either from DD simulation or calculated by MD model. We use this cost function for three reasons. First, the raw residual, $|J_{DD} - J_{MDB}|$, is included because this directly reflects how well the fitting result is. Second, given that the SRH recombination currents predominate in the cases we consider, the errors introduced by these currents, $|J_{DD}^{rec} - J_{MDB}^{rec}|$, should be taken into account, and other recombination currents are neglected. Third, the errors are weighted to the photocurrent, $J_{ph}$, such that the cost is normalized with respect to the

## WILEY-VCH

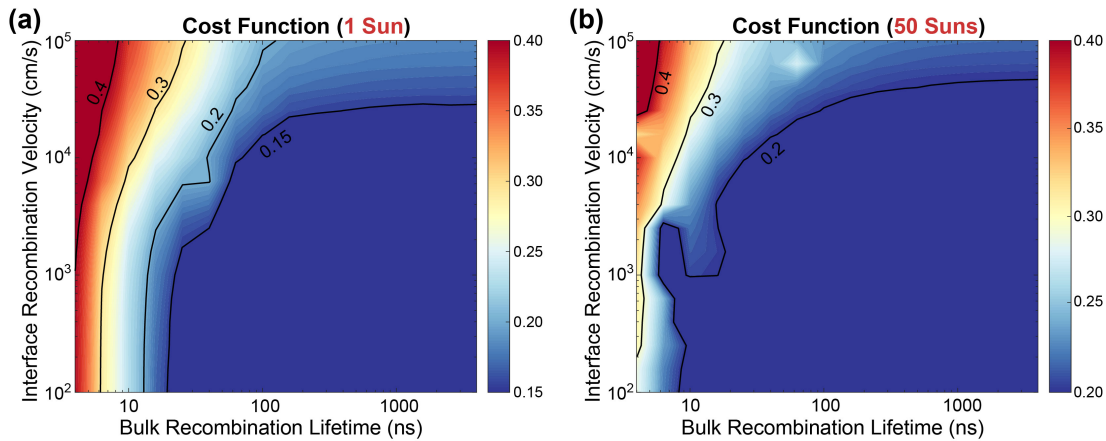

**Figure 4.** Contour plots of cost as a function of bulk SRH recombination lifetime and interface SRH recombination velocity for **(a)** 1 Sun irradiance and **(b)** 50 Suns irradiance. The cost function evaluates the accuracy of the MD model in describing the SRH recombination currents and the output current in comparison to DD simulations. A small value of the cost function indicates high accuracy.

irradiance. With these considerations, the cost function weighs the evaluation metric of major concern, i.e., how accurate the MD model describes the SRH recombination currents and output current in comparison to DD simulations: the smaller the cost function, the better the MD model performs.

In simulations, numerous parameters can be varied to produce $JV$ curves that exhibit different features. Out of these parameters, we mostly care about those that are directly related to recombination and resistances; therefore, we modify bulk SRH recombination lifetime, interface SRH recombination velocity, series resistance, and shunt resistance to evaluate the MD model via the cost function. To clearly visualize the cost function, we vary two simulation parameters simultaneously, avoiding high dimensional data. We perform the evaluation procedures as follows. First, starting from the device analyzed in the previous sections, simulation parameters are modified in broad ranges that cover most of the practical cases. For example, the bulk SRH recombination lifetime varies from nanoseconds to microseconds, corresponding to the bad and good bulk layers that can be possibly made in practice. Besides, parameters are varied in a nested way such that the interactive impact of both parameters on the cost function can be assessed. Second, each $JV$ curve simulated in the first step is then analyzed by the MD model, retrieving the values for lumped circuit parameters. Finally, using the retrieved parameters, we can calculate cost functions for each combination of simulation parameters.

Figure 4 illustrates the contour plot of cost as a function of bulk SRH recombination lifetime and interface SRH recombination velocity for devices under 1 Sun and 50 Suns illumination. The device parameters are consistent with those in Table S1, except that the perovskite layer thickness is 760 nm (close to the measured thickness of our lab-produced device in the next section). Having examined case by case, we notice that a cost function lower than 0.15 for 1 Sun irradiance and 0.2 for 50 Suns irradiance always indicates good fitting, where the predominant recombination current density falls in the close proximity of the simulated one in the exponential region. Therefore, these two values are chosen as the lower bounds of the cost function in Fig. 4, serving as the benchmark values to discern the goodness of fitting results. For reference, the cost function for the device in Fig. 1 is 0.02 under 1 Sun irradiance, and 0.06 under 50 Suns irradiance, representing very good fitting.

In Fig. 4(a), there are two noticeable trends worth discussing. 1) First, increasing the interface SRH recombination velocity to a very high level can lead to a slight increase in the cost function. This trend, evidenced by the fitting results in Fig. S3, is associated with the shunt resistance in the MD model. As mentioned previously, the shunt resistance takes on the role of describing recombination currents that are non-exponential in the low-voltage region. When SRH recombination is severe, the retrieved shunt resistance can be greatly reduced, but a shunt resistance with a fixed value cannot perfectly describe the variation in current density in the low-voltage region, thus introducing errors into the model. 2) The second trend is that diminishing bulk SRH recombination lifetime to the order of several nanoseconds increases the cost function dramatically. This trend is related to the way we use the MD model where the bulk SRH recombination current has a fixed coefficient and a fixed ideality factor. In a device with severe bulk SRH recombination, the open circuit voltage can drop below a certain level such that the voltage-injected carrier concentrations in the bulk cannot sustain charge neutrality. Therefore, using a fixed ideality factor of 2, which results from equal carrier concentrations, can make retrieved parameters deviate from their actual values, thereby increasing the cost function. In Fig. S4, we can see that for a device with a bulk SRH recombination lifetime of 10 ns, the bulk SRH recombination current calculated by the MD model does not closely follow the simulated one, thereby introducing errors in the exponential region. Besides, the shunt resistance accounts for part of the recombination current that further introduces errors in the low-voltage region, as in the case where interface SRH recombination is severe.

For the same reasons, similar trends can be observed for the same device under 50 Suns irradiance, as shown in Fig. 4(b), except that the variation in the cost function is less uniform and the range for good fitting is larger. Thus, it is possible to retrieve accurate parameters even with a very low bulk SRH recombination lifetime, so long as $JV$ curves under various irradiances (e.g., 50 Suns) are available. Besides, in Fig. S3 and S4, we notice that although the cost function is large in the case of severe SRH recombination, if we attribute the shunt current to the predominant recombination current, the loss analysis can be more accurate. Moreover, the accuracy of parameters retrieved under various irradiances can be different, which suggests a way of validating the accuracy of parameters by applying the MD model to the same device under various irradiances: if parameters retrieved under different irradiances are close or equal, then it is more likely that the fitting results are accurate.

Next, we study the impact of resistances on the cost function. Figure S5(a-b) illustrates the contour plot of cost as a function of series resistance and shunt resistance for devices under 1 Sun and 50 Suns illumination. Notably, the cost function at 1 Sun irradiance is uniformly lower than 0.03, indicating that the accuracy of the MD model is unrelated to the values of resistances. Under 50 Suns illumination, however, cost functions with high values (around 0.9) exhibit a patterned distribution when series resistance is large. These outliers, examined by individual fitting, turn out to be resultant from inappropriate initial guesses. In Fig. S5(c), we apply the MD model using two different sets of initial guesses for the same device that has a series resistance of $10 \ \Omega \cdot cm^2$ and a shunt resistance of $1000 \ \Omega \cdot cm^2$ under 50 Suns irradiance; the cost function is 0.896 for the first set of initial guesses with apparently inaccurate fitting results, while the cost function is 0.039 for the other set of initial guesses with very accurate fitting results. Such a phenomenon occurs because the algorithm can get stuck at the local minimum, suggesting that a randomized set of initial guesses can be used in order to avoid the local minimum and achieve good fitting. This feature can be easily incorporated into the model via several times of fitting with randomized initial guesses, in which the fitting result with the least residual can be selected as the final result. Furthermore, since irradiance may affect the accuracy of fitting, we can use the parameters retrieved from a specific irradiance that is accurate as the initial guess for fitting under other irradiances. We adopt this strategy by using the series resistance retrieved under 1 Sun as the initial guess for fitting under 50 Suns. The result is shown in Fig. S5(d), where the cost function for all resistance values under consideration is uniformly below 0.07. Therefore, we can conclude that the values of resistances do not have an impact on the accuracy of the MD model.

In sum, the MD model can discern and quantify different loss pathways in most of the practical cases, whose accuracy is associated with both interface and bulk SRH recombination. Two special cases, where fitting results can be erroneous, require further interpretation. First, in the case where bulk SRH recombination lifetime is on the order of several nanoseconds or worse, the fitting errors can be significant, and we should directly conclude from the TRPL results that the bulk of the absorption layer needs to be optimized first. Second, when interface SRH recombination is severe, with a good bulk layer and negligible current leakage, a large fitting error in the low-voltage region may be introduced because the usage of a fixed $R_{sh}$ that in part takes on the role of describing recombination currents. In this case, we should attribute the loss due to shunt resistance in the MD model to interface SRH recombination so that the loss analysis can be more accurate. With the knowledge gained from the above analysis, in the following section, we apply the MD model to our lab-produced devices and quantify the corresponding losses in each case. The results give us useful information on the losses in our devices, which in turn validates the effectiveness of this approach.

## 3. Experiments

Before analyzing our experimental results, it is necessary to discuss the ion migration that gives rise to hysteresis in many perovskite devices. The mobile ions in the perovskite absorber can lead to an extra complication in interpreting the results obtained by the MD model. Although the well-established DD simulator (SCAPS-1D) is widely used in the literature to study perovskite photovoltaics[7,14,36–38], it does not incorporate the mobile ions in the perovskite layer. To investigate the performance of the MD model in the presence of mobile ions, we simulated two sets of steady-state $JV$ curves incorporating mobile ions with SolarDesign[39] and IonMonger[40]. The reason for studying steady-state $JV$ curves is that the effect of ions in steady states is unchanged and the practical performance of photovoltaics is evaluated at steady-state maximum power points. The device parameters and simulation details are given in Supplementary Note 2. In the first set of results, bulk SRH recombination is the predominant loss pathway. As shown in Fig. S6, the impact of ions on the steady-state $JV$ curves has a negative correlation with the bulk SRH recombination lifetime, which is consistent with the trend in Fig. 4, suggesting the broad applicability of the MD model. However, the perceptible differences, though being small for large bulk SRH recombination lifetime, raise concern regarding the accuracy of describing bulk SRH recombination by Eq. (2), which is a function of QFLS and relies on charge neutrality. Accounting for mobile ions, the electric field in the device can be modified by an inhomogeneous distribution of ionic charges, potentially affecting charge neutrality via the redistribution of electrons and holes. Fortunately, the ionic charges tend to accumulate at the interfaces, resulting in the field screening effect[41,42] such that the carriers inside distribute even more freely, thereby enhancing charge neutrality. Figure S7 illustrates the ionic and electronic distributions and the corresponding band diagrams, which can be used to analyze device performance in the presence of ions. At thermal equilibrium, the ions (and the vacancies) accumulate at the perovskite-transport-layer interfaces, resulting in band bending at the interfaces and band flattening in the perovskite bulk. From the perspective of classical pn-junction theory[43], the accumulated charges at the interfaces form two depletion regions. Upon illumination, the depletion regions shrink immediately and part of the ionic charges are screened by the carriers, which will then diffuse into the perovskite bulk. This ionic diffusion process semibounds the carriers, leading to a decrease in transient currents. After sufficient illumination time, the ionic and electronic charges will reach a steady state with fewer interfacial ions. Under high irradiances or large applied voltages, the electronic charges prevail in the device, and the impact of ion accumulation on carrier distributions in steady states is thus smaller[43]. Moreover, the MD model captures the dynamics of recombination currents in the exponential region (where the applied voltage is large), suggesting that the MD model can be applied to steady-state $JV$ curves effectively. As shown in Fig. S8(a), the predominant recombination current computed by the MD model can follow the simulated result in the exponential region. Such consistency suggests that in the presence of ions (accumulating at the interfaces), QLFS still has the exponential form of $q(V + JR_s)$, which is confirmed by the energy level diagrams shown in Fig. S8(c-d) and the charge distribution diagrams shown in Fig. S9. The second set of results is simulated for the device with predominant interface SRH recombination. As shown in Fig. S10, the impact of ions on the steady-state $JV$ curves depends on the SRH recombination velocities at the interfaces. The ions may increase or decrease the steady-state performance, depending on the properties of transport layers[44]. In both cases, the MD model can accurately capture the predominant loss current (as shown in Fig. S8(b)), because Eq. (3) has a variable ideality factor. Therefore, the trend in Fig. 4 can be used as the accuracy indicator for the MD model applying to steady-state $JV$ curves, showing where the losses can be analyzed effectively. Moreover, with the development of fabrication technology, perovskite photovoltaics that are less affected by ions[45] are expected to lead future research and promote the commercialization of perovskite photovoltaics[46].

To explore the capability of the MD model in practice, we apply it to two types of devices that both use $Cs_{0.05}FA_{0.95}PbI_3$ as the active layer composition, but are made in different conditions. The perovskite layer of Device 1 was prepared in ambient air with a humidity of 60%, while Device 2 was made in the $N_2$ glove box with standard procedure. Both devices are of pin type shown in Fig. S11 and the fabrication details are specified in Supplementary Note 5. Figure 5 depicts the measured $JV$ curves for both devices under AM 1.5G illumination with a sampling rate of 5 NPLCs and 100 data points per scan (NPLC, number of power line cycles, is the unit that Keithley source meters use to specify scanning rates; the sampling method we adopted corresponds to a scanning rate of ~0.1 V/s). Noticeably, the $JV$ curves only show negligible hysteresis, indicating that the operating points are close to their steady states. To verify this, we additionally measured a set of steady-state current density values by holding the

## WILEY-VCH

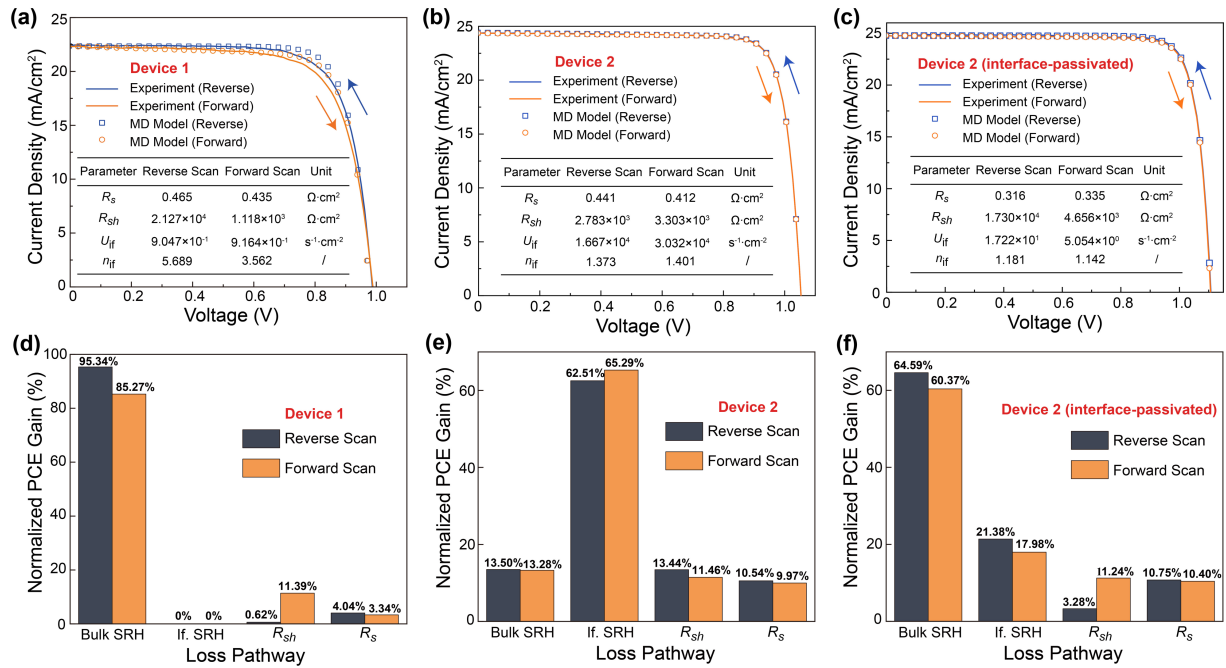

**Figure 5. (a-c)** Current density-voltage relations of our lab-produced perovskite photovoltaics (Device 1, Device 2, and Device 2 with passivated interface). The MD model fitted $JV$ curves are also depicted in the corresponding figures, with retrieved parameters listed in the inset table for both forward- and reverse-scan $JV$ curves. **(d-f)** Normalized PCE gains for each loss pathway, calculated based on the retrieved MD model parameters.

voltage at fixed levels for ~60 seconds and sampling at the same rate of 5 NPLCs. The range and average values of the recorded steady-state data points are given in Fig. S12, which are in close proximity to the forward and reverse $JV$ curves. Also, for both devices during steady-state measurement, we notice that the current density only fluctuates within a small range, rather than shifting towards higher or lower values. Therefore, the forward and reverse $JV$ curves are close to the steady states, and in what follows, we apply the MD model to both forward and reverse $JV$ curves, analyzing the underlying losses in these devices.

In order to describe the bulk SRH recombination current in Eq. (2), we measure the active layer thickness, the bulk SRH recombination lifetime (for describing $\gamma_{bulk}$), and the bandgap of the active layer (for estimating $n_i$). The active layer thickness is ~750 nm for both devices, measured by a step profiler (Supplementary Note 5). As shown in Fig. S13, the bandgaps of perovskite in both devices are ~1.55 eV, measured by an ultraviolet-visible spectrometer. From TRPL results (Fig. S14), the bulk SRH recombination lifetime is 21 ns for Device 1, and 398 ns for Device 2. With all these data, we apply the MD model to both devices, retrieving the values for lumped parameters, then calculating the corresponding $JV$ curves and quantifying losses in terms of normalized PCE gain shown in Fig. 5.

Since in terms of recombination lifetime, Device 2 has a ~20 times better perovskite bulk layer than Device 1, it is expected that the analysis via the MD model can lead to some differences in loss quantification. As can be observed, in Fig. 5(a), the fitted $JV$ curves for Device 1 deviate slightly from the measured ones at the voltages where the exponential tail starts, and Fig. 5(d) shows that the bulk SRH recombination is the predominant loss pathway in terms of normalized PCE gain. On the contrary, the fitting errors are extremely small for Device 2 in both scan directions, and the loss quantification in Fig. 5(e) shows that the interface SRH recombination loss dominates. These results are consistent with the simulation results in Fig. 4, where devices with severer bulk SRH recombination can have greater fitting errors. The results also show that the differences in recombination lifetime can be successfully translated into the differences in loss quantification via the MD model, indicating that the MD model is capable of discerning and quantifying losses in perovskite photovoltaics in practice. Moreover, the MD model provides us with the guidance on optimizing the device: the PCE of Device 1 can be greatly improved by diminishing the bulk SRH recombination, which is already accounted for in Device 2; on the other hand, to improve the PCE of Device 2, the most effective strategy is to passivate the interfaces. Following this guidance, we passivate the perovskite/BCP interface of Device 2 with n-octylammonium iodide (OAI), to further examine the performance of the MD model. The $JV$ curves of the interface-passivated Device 2 are shown in Fig. 5(c), which, compared with those of non-passivated Device 2, have a slight increase in open-circuit voltage (~0.05 V) and short-circuit current density (~0.4 mA/cm$^2$). In Fig. 5(f), these slight differences translate into a significant boost of bulk SRH recombination loss and a decrease in interface SRH recombination loss, meaning that the predominant loss pathway shifts from interface SRH recombination to bulk SRH recombination. This phenomenon reveals that using (OAI) is an effective way of passivating the perovskite/BCP interface if the MD model is accurate, and since the strategy of passivation using organic onium salts is validated in many literatures[47–49], this in turn corroborates the effectiveness of the MD model applied in practical devices.

Next, to evaluate the high irradiance performance of the MD model, we measure the $JV$ curves for Device 2 under irradiances up to 5 W/cm$^2$ (50 Suns), and apply the MD model to these $JV$ curves. The light source is a 760 nm diode laser, whose power is measured by an integrating sphere photodiode power sensor (Thorlabs S142C), and the sampling rate is still fixed at 5 NPLCs. The fitting results are shown in Fig. 6(a-f) and the retrieved lumped parameter values are listed in Table S2. It is significant that the fitting results are quite accurate for all these irradiance levels, which strongly indicates that the MD model captures the behaviors of

**WILEY**-VCH

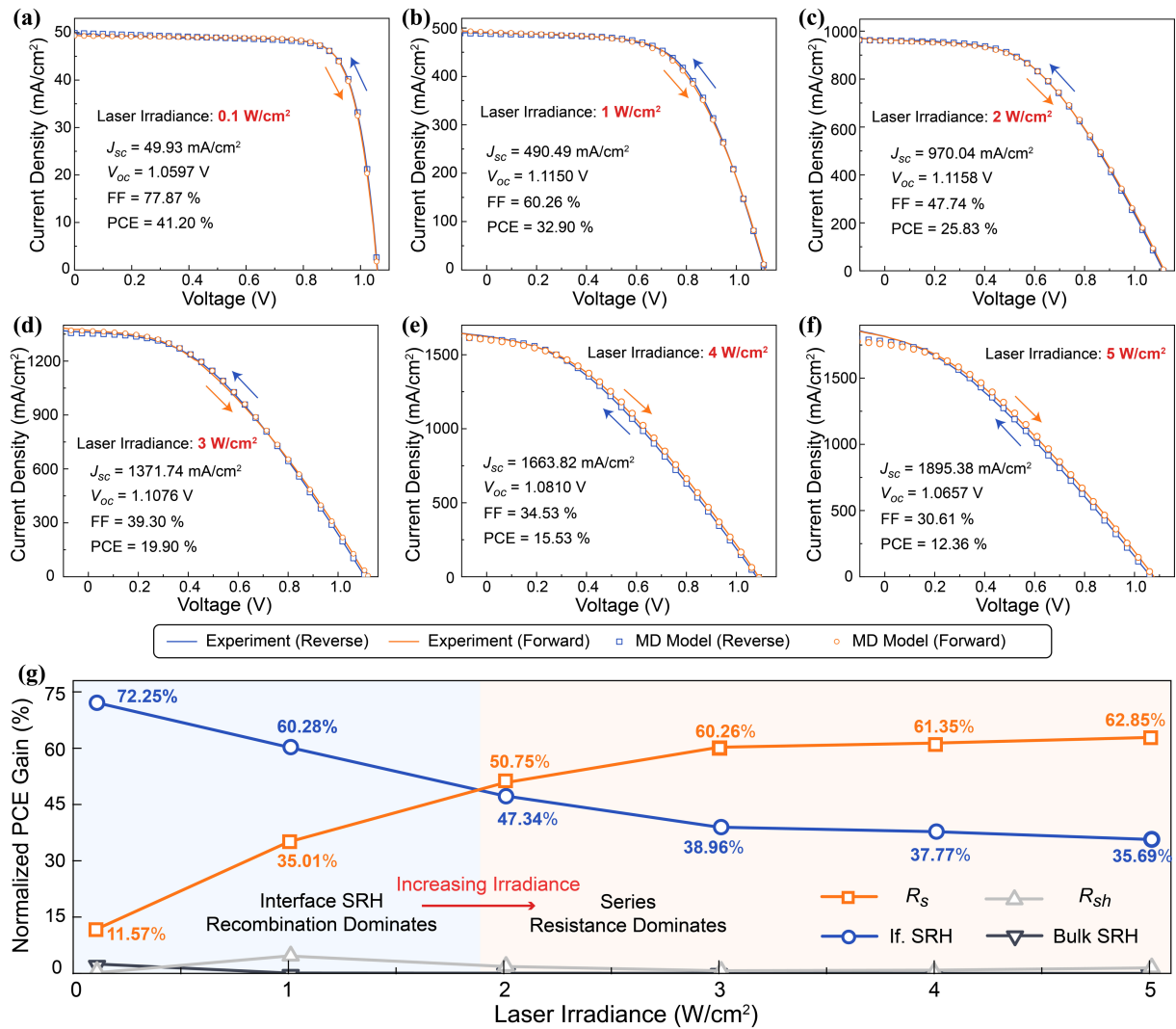

**Figure 6. (a-f)** Current density-voltage relations of Device 2 under laser (760 nm) irradiances from 0.1 W/cm² to 5 W/cm². The MD model fitted $JV$ curves are also depicted in the corresponding figures. **(g)** The normalized PCE gain of each loss pathway as a function of laser irradiance.

the devices under high irradiances and ionic effects are indeed negligible in this device. Besides, we notice that: 1) The retrieved values for series resistance are of the same level ($\sim 0.4\ \Omega \cdot \text{cm}^2$). As discussed previously, such consistency indicates the accuracy of the fitting results. 2) Fitting errors at short circuit are slightly growing with increasing irradiance. In line with the simulation results shown in Fig. S3, these growing errors are introduced due to the usage of a fixed $R_{sh}$ in the MD model, which in part takes on the role of describing recombination currents in the low-voltage region. With increasing irradiance, the recombination currents in the low-voltage region become more pronounced, and thus, a fixed $R_{sh}$ cannot fully capture such dynamics. This inconsistency also corroborates the fact that, in this device, the current leakage is negligible and the loss due to $R_{sh}$ in the MD model should be attributed to the predominant recombination loss. Figure 6(g) shows the normalized PCE gains for each loss pathway at different irradiance levels. The losses due to interface SRH recombination and series resistance are the two major competing loss pathways in this device. Below $2\,\mathrm{W/cm^2}$ (20 Suns), the interface SRH recombination is the predominant loss; with increasing irradiance, the series resistance gradually becomes predominant. This is consistent with the results in the simulation study shown in Fig. 2, and the reason why series resistance loss becomes predominant can be understood with the MD model from another perspective: in terms of power density, the loss due to $R_s$ can be directly calculated by $J^2 R_s$ as per Joule's law, where the current density, $J$, approximately scales linearly with irradiance; therefore, the power loss due to $R_s$ has a quadratic growth rate with respect to irradiance, which will take a larger proportion under higher irradiances.

To sum up, in this section, the experimental study demonstrates that the MD model is capable of discerning and quantifying electrical losses in perovskite devices, for both 1 Sun and high irradiance applications. Notably, the details of fitting results are consistent with those in the simulation studies, which further validates the effectiveness of this approach.

## 4. Guidelines for Applying the MD Model

For reference, we summarize the procedures that we use to analyze the devices in this paper as a set of guidelines. With the following steps, one can use the MD model to quantify and analyze the losses in perovskite photovoltaics:

1) **Measure the $JV$ curve of the device.** The MD model is expected to be performant when applied to steady-state $JV$ curves even in the presence of mobile ions. Thus, the scanning rate during measurement should be low enough for the completion of ionic motions, or a set of stabilized currents at fixed voltages (SCFV) can be measured to directly produce the steady-state $JV$ curve[50].

2) **Measure the necessary parameters for describing bulk SRH recombination.** To properly describe the bulk SRH recombination current in Eq. (2), the active layer thickness, the bulk SRH recombination lifetime (for describing $\gamma_{bulk}$), and the active layer bandgap (for estimating $n_i$) need to be measured.

3) **Apply the MD model to the JV curves measured in Step 1 with the parameters measured in Step 2.** For convenience, one can directly use our MATLAB implementation to apply the MD model, which is open-source online[31].

4) **Quantify the losses in terms of normalized PCE gain.** The effect of a loss mechanism in a complete device under operational conditions can be quantified by a specific indicator such as the normalized PCE gain proposed in this paper. One can also devise other indicators, so long as the dependence among different losses is taken into account. After pinpointing the predominant loss pathway, one can refer to Table S4 to devise corresponding optimization strategies.

5) **Interpret the results if the fitting error is large.** There are several cases that can cause a large fitting error. First, if the bulk SRH recombination lifetime is on the order of several nanoseconds or worse, the fitting error may be large; in this case, we should conclude from the TRPL results that the bulk of the absorption layer needs to be optimized first. Second, in the case of severe interface SRH recombination or high irradiance, with a good bulk layer and negligible current leakage, a large fitting error in the low-voltage region may be introduced because the usage of a fixed $R_{sh}$ that in part takes on the role of describing recombination currents; despite the error, the fitting results in the exponential region are accurate (as in Fig. 6), so the retrieved parameters can still be used to analyze losses. Third, a large error can be introduced if the algorithm used by the MD model gets stuck at local minimums with inappropriate initial conditions. One can use a randomized set of initial guesses to obtain the results with the minimum error. The causes of large fitting error and the corresponding solutions are summaried in Tabls S5.

## 5. Future Research

As a circuit model that can accurately quantify electrical losses, the MD model can be generalized or included in a larger framework, in order to solve different problems. Here, we list three potential research topics where the MD model can play an important role:

1) The MD model can be coupled with an optical model such as the transfer matrix method to extend the analysis that includes optical loss. Since the MD model can accurately quantify electrical losses, the accuracy of the overall loss analysis can be improved.

2) The MD model itself can be generalized to other photovoltaic technologies where two competing recombination processes are entangled, so long as the characteristic of one process can be identified by experimental techniques.

3) The MD model can be applied as the elementary component in a distributed circuit model[51,52] so as to analyze two-dimensional effects, such as non-uniform illumination, distributed resistance, and so forth.

## 6. Conclusion

In this work, the MD is proposed to quantify bulk/interface defect-assisted recombination and series/shunt resistive losses. Combined simulation and experimental studies show that in most practical cases, the MD model can accurately quantify all the aforementioned losses, and in some special cases including severe bulk or interface SRH recombination, it is possible to pinpoint the predominant loss pathway. Moreover, at higher irradiances, the MD model also

## WILEY-VCH

demonstrates good performance with high accuracy in both simulation and experimental studies, which makes it feasible for evaluating photovoltaics in applications such as concentrators and optical wireless power transfer. Finally, we provide a set of guidelines for applying the MD model and interpreting the results. We make the source code freely available online, which will hopefully facilitate the design of efficiency-targeted optimization strategies for researchers in this field.

## Supporting Information

Supporting Information is available from xxx

## Acknowledgements

Minshen Lin and Xuehui Xu contributed equally to this work.

WILEY-VCH

## Supporting Information

## Quantifying Nonradiative Recombination and Resistive Losses in Perovskite Photovoltaics: A Modified Diode Model Approach

*Minshen Lin[1], Xuehui Xu[2], Hong Tian[2], Yang (Michael) Yang[2]\*, Wei E. I. Sha[3]\*, and Wenxing Zhong[1]\**

[1]College of Electrical Engineering, Zhejiang University, Hangzhou 310027, China.

[2]State Key Laboratory of Modern Optical Instrumentation, Institute for Advanced Photonics, College of Optical Science and Engineering, Zhejiang University, Hangzhou 310027, China

[3]State Key Laboratory of Modern Optical Instrumentation, College of Information Science and Electronic Engineering, Zhejiang University, Hangzhou 310027, China

*Authors to whom correspondence should be addressed: yangyang15@zju.edu.cn, weisha@zju.edu.cn, and wxzhong@zju.edu.cn

### Note 1. Defect-assisted recombination current from Shockley-Read-Hall model

The Shockley-Read-Hall model[1] describes the defect-assisted recombination rate as

$$U_{SRH} \approx \frac{np - n_i^2}{\tau_p n + \tau_n p}, \tag{S1}$$

where $\tau_n$ and $\tau_p$ are the respective defect-assisted recombination lifetimes for electrons and holes, inversely proportional to the concentration of defect states. As discussed in Section II, with higher applied voltage, more carriers are electrically injected into the bulk of the device and charge neutrality can gradually establish. This process is also illustrated in Fig. S1, where a device with predominant bulk SRH recombination is analyzed in terms of energy level and carrier concentration distributions. Moreover, the quasi-Fermi level splitting (QFLS) is equal to $q(V + JR_s)$ in the MD model, and $np = n_i^2 \exp(\text{QFLS}/k_B T)$. In this case ($n \approx p$), Eq. (S1) can be simplified to

$$U_{SRH} = \frac{np - n_i^2}{(\tau_p + \tau_n)\sqrt{np}} = \gamma_{bulk} n_i \left[ \exp\left(\frac{V + JR_s}{2k_B T/q}\right) - 1 \right], \tag{S2}$$

where $\gamma_{bulk} = 1/(\tau_p + \tau_n)$ is the bulk SRH recombination coefficient. Integrating over the length of the absorption layer, $L$, we have the bulk SRH recombination current density,

$$J_{bulk}^{SRH} = qU_{SRH}L = qL\gamma_{bulk}n_i\left[\exp\left(\frac{V+JR_s}{2k_BT/q}\right)-1\right]. \tag{S3}$$

Likewise, the interface SRH recombination at the transport layer/perovskite interface can be described by the SRH formulation as

$$U_{SRH}^{if} \approx \frac{n^-p^+ - n_i^-n_i^+}{n^-/S_p + p^+/S_n}, \tag{S4}$$

where the superscripts of $\pm$ denote quantities evaluated at either the transport layer or perovskite side of the interfaces, and $S_n$, $S_p$ are electron and hole recombination velocities[2]. However, at the interfaces, there is no simple relation between electron and hole densities that can be captured for simplifying Eq. (S4); hence, we use Eq. (3) with two undetermined parameters to describe the interface recombination current via curve fitting.

**Note 2. Drift diffusion simulations**

To study the performance of the modified diode (MD) model, we use drift-diffusion (DD) simulation results as the benchmarks for comparison. We use the well-established DD simulator, SCAPS[3], to produce the simulation results without regard to mobile ions. Unless particularly specified, the devices in DD simulations have the parameter values listed in Table S1. Besides, in the DD simulations, the radiative recombination coefficient is determined in such a way that the volumetric radiative recombination current equals the areal radiative recombination current calculated by the principle of detailed balance. This setting fulfills the principle of detail balance and in the limiting case, the DD simulations can reproduce the detailed balance limits[4]. In the course of curve fitting, all the parameters in the MD model are allowed to vary within a broad range (from 0 to $10^{12}$).

To investigate the impact of ions on the steady-state performance of perovskite photovoltaics, SolarDesign[2] and IonMonger[3] are used to simulate the $JV$ curves with varying ionic vacancy densities. The device parameters used for simulation are mostly the same as the template device[3], except for the parameters associated with bulk and interface SRH recombination, which are given in Table S3.

The details of modeling methods are described as follows. The governing equations of the drift-diffusion model without regarding to mobile ions are:

$$\begin{cases} \nabla \cdot (\varepsilon_r \nabla \psi) = q(n-p), \\ \dfrac{\partial n}{\partial t} = \dfrac{1}{q}\nabla \cdot J_n + G - R, \\ \dfrac{\partial p}{\partial t} = -\dfrac{1}{q}\nabla \cdot J_p + G - R, \end{cases} \tag{S10}$$

## WILEY-VCH

where $J_n = -q\mu_n n \nabla\psi + qD_n\nabla n$ and $J_p = -q\mu_p p\nabla\psi - qD_p\nabla p$ are the electron and hole current densities, respectively. The electron (hole) diffusion coefficient satisfies the Einstein relation $D_{n(p)} = \mu_{n(p)}k_BT/q$ and $\mu_{n(p)}$ is the electron (hole) mobility. Furthermore, $G = G^{ph} + G^{dark}$ is the total generation rate, where $G^{ph}$ is the photon generation and $G^{dark}$ is the dark generation at thermal equilibrium. Similarly, $R = R^{rad} + R^{nonrad}$ is the recombination rate at non-equilibrium states where $R^{rad}$ is the radiative recombination rate and $R^{nonrad}$ is the non-radiative recombination rate.

In perovskite photovoltaics, the non-radiative recombination is mainly due to the following defect-assisted Shockley-Read-Hall (SRH) recombination. The bulk SRH recombination rate is given by

$$R_{SRH} = \frac{np}{\tau_n(p + p_t) + \tau_p(n + n_t)}, \tag{S11}$$

where $\tau_n$ and $\tau_p$ are the lifetimes of excess electrons and holes, respectively. The trap level $E_t$ in the bandgap is used to compute the densities of electrons and holes with respect to the trap level, i.e., $n_t = N_C\exp[(E_t - E_C)/k_BT]$ and $p_t = N_V\exp[(E_V - E_t)/k_BT]$. Similarly, the SRH recombination fluxes $R_{SRH}^E$ and $R_{SRH}^H$ at the interfaces are given by[3]

$$R_{SRH}^{E,H} = \frac{n^- p^+}{\frac{1}{v_n^{E,H}}(p^+ + p_t^+) + \frac{1}{v_p^{E,H}}(n^- + n_t^-)}, \tag{S11}$$

where the superscripts $\pm$ denote quantifies evaluated at either the left- or right-hand side of the perovskite/transport layer interfaces, respectively.

To account for mobile ions in the perovskite layer, in Eq. (S10), Poisson's equation needs to be modified to include the immobile cation vacancies $N_0$ and halide ion vacancy density $P$; also, a continuous equation for ions needs to be added.

$$\begin{cases} \nabla \cdot (\varepsilon_r\nabla\psi) = q(n - p + N_0 - P), \\ \dfrac{\partial n}{\partial t} = \dfrac{1}{q}\nabla \cdot J_n + G - R, \\ \dfrac{\partial p}{\partial t} = -\dfrac{1}{q}\nabla \cdot J_p + G - R, \\ \dfrac{\partial P}{\partial t} = \nabla \cdot F^P, \end{cases} \tag{S12}$$

where $F^P = D_I\nabla P + \mu_I P\nabla\psi$ is the flux of ion vacancy density, and $D_I$ and $\mu_I$ are the ionic diffusion coefficient and mobility, respectively.

**Note 3. Loss analysis**

**WILEY-VCH**

To study the impact of each loss pathway on the power conversion efficiency (PCE) of photovoltaics, we introduce an approach to calculating the normalized PCE gains. As discussed in Section 2.2, the influences of each loss are coupled in such a way that reducing one loss can give rise to a decrease in the others. We choose the maximum power point (MPP) of the device, where the PCE is evaluated, as the point of interest for comparison. Based on the device involving all losses at hand, in the DD simulation or the MD model, we exclude each loss pathway one at a time and calculate the PCE gains with respect to each loss. In the DD simulation, we exclude the loss pathway by making the corresponding simulation parameter inactive, while in the MD model, we do this by deleting the corresponding term in Eq. (1): for example, we exclude $R_s$ by setting the value of $R_s$ to zero yet keeping others the same, or we just delete the term $J_{SRH}^{bulk}$ to exclude the effect of bulk SRH recombination. In this way, we can evaluate the impact of each loss, and normalize them as the normalized PCE gain:

$$\text{Normalized PCE Gain} = \frac{\text{PCE Gain without Loss } i}{\sum_i \text{PCE Gain without Loss } i} \times 100\%. \tag{S5}$$

This indicator can be viewed as a quantity that describes the influence of a loss pathway in percentage, which demonstrates their relative importance in a complete device.

**Note 4. Differential ideality factor in the presence of series resistance**

Taking series resistance into account, the recombination current is given by

$$J_{rec} = J_o \exp\left[\frac{q(V + JR_s)}{n_{id}k_BT}\right]. \tag{S6}$$

Assuming that the ideality factor, $n_{id}$, and the series resistance, $R_s$, are independent of applied voltage (this is consistent with the parameter retrieval process of curve fitting that produces voltage-independent parameters), we can find an expression for $n_{id}$ by differentiating the above equation with respect to $V$, i.e.,

$$\frac{dJ_{rec}}{dV} = J_{rec}\frac{q}{n_{id}k_BT}\left(1 + \frac{dJ}{dV}R_s\right). \tag{S7}$$

Reordering the terms, we have

$$n_{id} = \frac{q}{k_BT}\left(\frac{d\ln J_{rec}}{dV}\right)^{-1}\left(1 + \frac{dJ}{dV}R_s\right). \tag{S8}$$

In the exponential region of the $JV$ curve, $J_{rec}$ is the predominant current density loss such that $dJ/dV = -dJ_{rec}/dV$. Thus, we can substitute $J_{rec}$ in place of $J$,

$$n_{id} = \frac{q}{k_BT}\left(\frac{d\ln J_{rec}}{dV}\right)^{-1}\left(1 - \frac{dJ_{rec}}{dV}R_s\right). \tag{S9}$$

Using Eq. (S9), the calculated $n_{id}$ varies with voltage, and the retrieved value lies in the exponential region, close to the minimum differential ideality factor, as shown in Fig. 3.

**Note 5. Device Fabrication and Characterization**

The devices in this paper were made from commercially available products. $C_{60}$, FAI, $PbI_2$, CsI, n-Octylammonium Iodide (OAI), and BCP were purchased from Xi'an Polymer Light Technology Corporation. Dimethyl formamide (DMF), dimethyl sulfoxide (DMSO), acetone, and chlorobenzene (CB) isopropanol (IPA), were purchased from J&K Scientific. Meo-2PACz were purchased from TCI(Shanghai)Development Co., Ltd. ITO substrates (86% transmittance, 15 Ohm·sq$^{-1}$) were purchased from South China Xiangcheng Technology Co., Ltd.

The ITO glasses were sequentially cleaned in deionized water, acetone, and IPA by sonication for 5 min. Before deposition of the hole transporting layer, all ITO glasses were further cleaned for 20 min by a UV-ozone machine. To deposit the hole transporting layer, 0.4mg/ml Meo-2PACz iso-propyl alcohol solution was spin-coated onto ITO glasses at 5000 rpm for 30 s in the $N_2$ glove box and annealed at 100 °C for 10 min. The HTL thickness was measured to be 10 nm. The perovskite precursor solution was prepared by mixing 645.5 mg $PbI_2$, 228.7 mg FAI, 28.3 mg CsI, and 200 μL DMSO in 800 μL DMF solvent. All the solutions were stirred at 60 °C over 2 hours and filtered with a 0.22 μm polytetrafluoroethylene filter before use. The perovskite layer of Device 1 was prepared in ambient air with a humidity of 60%, while Device 2 was made in the $N_2$ glove box with standard procedure. The perovskite films with thicknesses of 750 nm were deposited by two-step spin-coating progress at first 1000 rpm for 10 s with a ramp of 1000 rpm·s$^{-1}$, and then 4000 rpm for 25 s with a ramp of 4000 rpm·s$^{-1}$. 200 μL chlorobenzene (CB) was dropped onto the film ~10 s before the end of the procedure and then annealed at 100 °C for 10 min under a different atmosphere. For the interface-passivated sample, 50ul of OAI (1mg/ml in IPA) was spun onto the perovskite layer with a speed of 5,000 rpm for 30 s without annealing. After that, 25nm $C_{60}$, 5nm BCP, and 100 nm Ag were coated onto the prepared sample by the thermal evaporation method sequentially.

The standard current density-voltage ($JV$) measurements and stabilized current at fixed voltage measurements were carried out by a Keithley 2400 source meter under AM 1.5G illumination from the xenon arc lamp of a Class A solar simulator. The light intensity was calibrated by a reference mono-crystalline Si solar cell. The $JV$ measurements under laser irradiance were carried out by a Keithley 2460 source meter under the illumination of a 760 nm laser. The irradiance of the laser was measured by an integrating sphere photodiode power sensor (Thorlabs S142C). The ultraviolet-visible (UV-Vis) absorbance measurements were conducted with an Agilent Cary 7000 UV-Vis spectrometer. The thicknesses of the films were

**WILEY-VCH**

measured by XF-WSBX-2200125 Step-Profiler. Time-resolved photoluminescence (TRPL) spectra were measured with a home-setup confocal fluorescence system at room temperature by a time-correlated single photon counting (TCSPC) module (PicoHarp 300), a scratch pad memory (SPAD) detector (IDQ, id100) with an instrument response function of ~100 ps, and a picosecond 532 nm laser.

**WILEY**-VCH

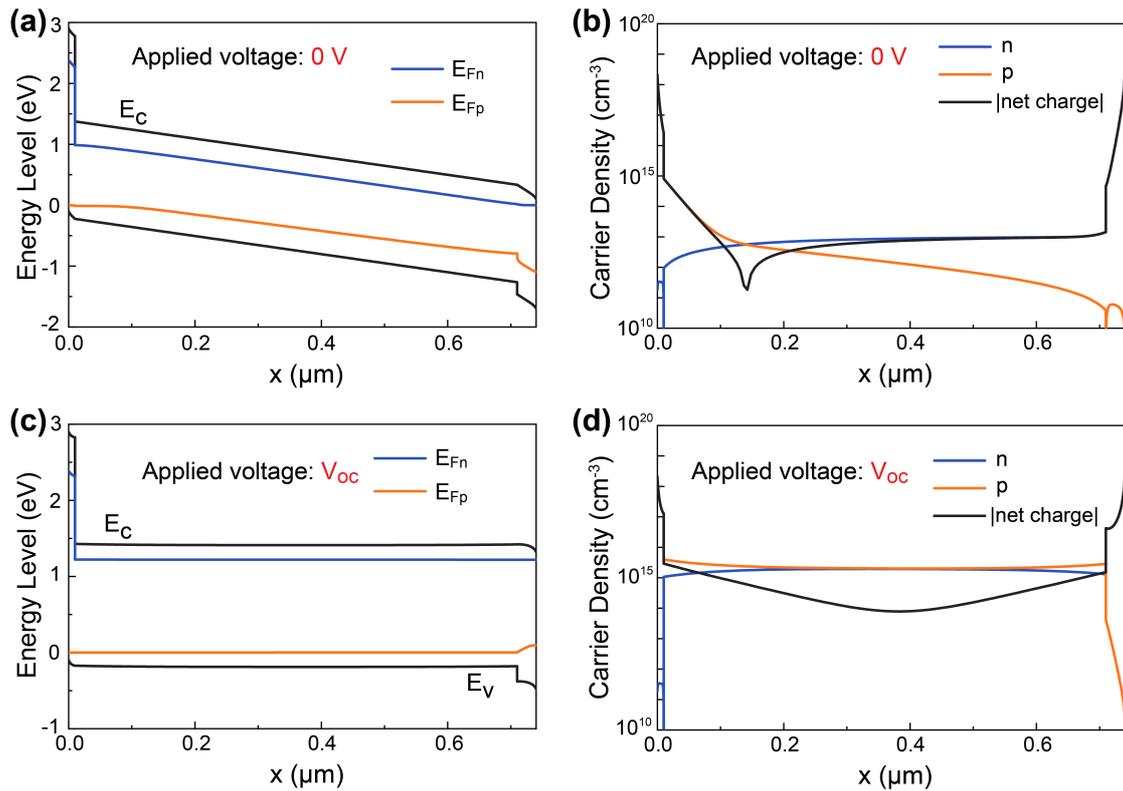

**Figure S1.** Energy level and carrier density distributions in the device with predominant bulk SRH recombination. **(a-b)** The distributions under an applied voltage of 0 V. **(c-d)** The distributions at open-circuit voltage. The irradiance is 1 Sun for all cases. It is noticeable in **(b)** and **(d)** that with higher applied voltage, more carriers are injected into the bulk of the device and charge neutrality establishes.

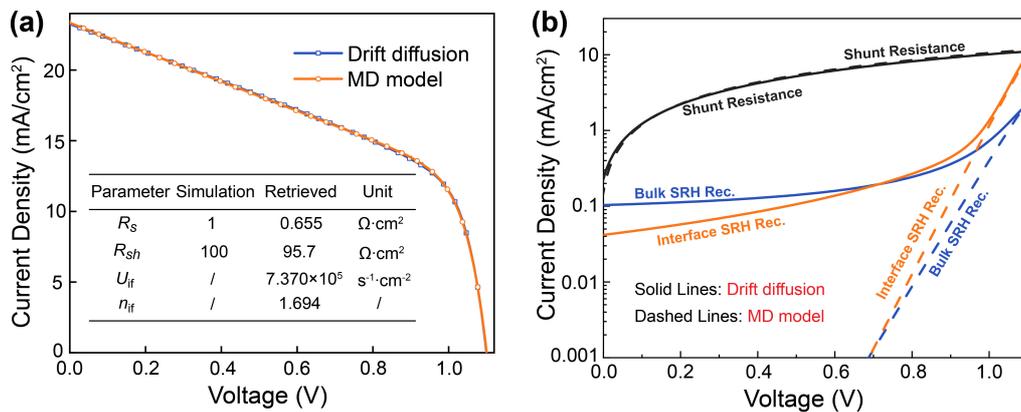

**Figure S2.** Current density-voltage relations of a device with severe shunt current leakage. The shunt current density calculated by MD model is consistent with the simulation result.

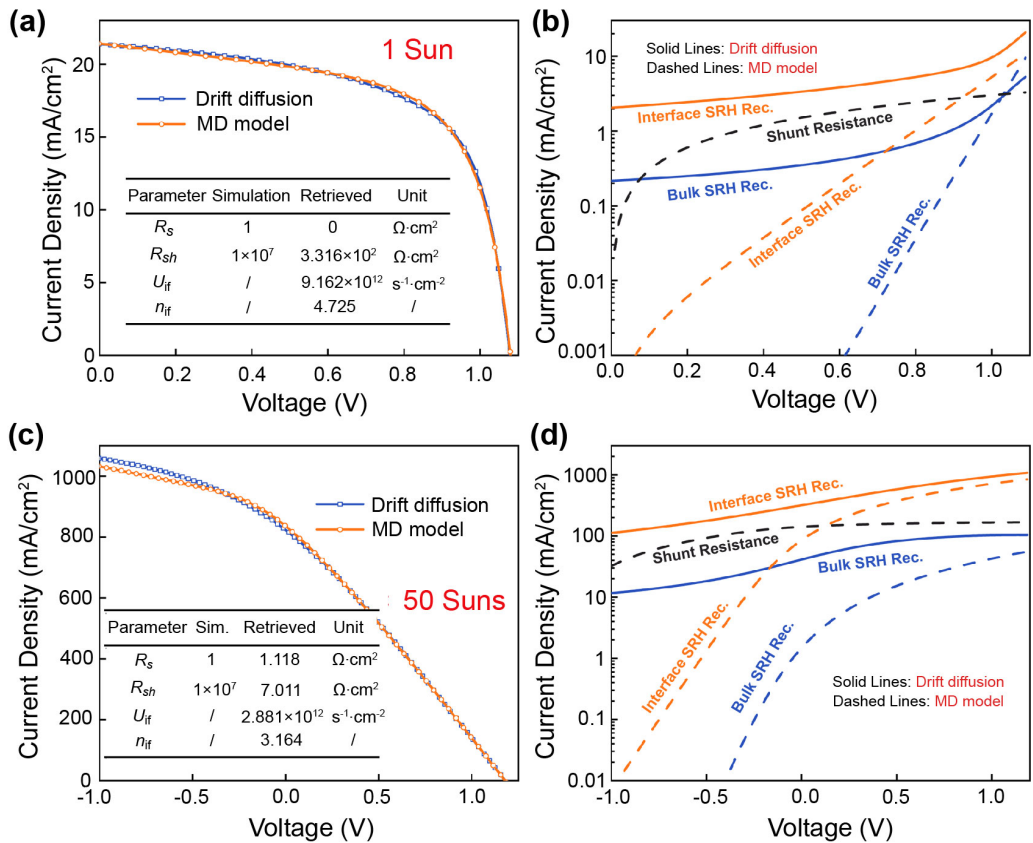

**Figure S3.** Current density-voltage relations of a device with severe interface SRH recombination. It is noticeable that the large recombination current in the low-voltage region bends the *JV* curve in a similar way as with $R_{sh}$. However, such increase in recombination current with respect to applied voltage cannot be perfectly described by $R_{sh}$ with a fixed value, thereby introducing errors into the cost function.

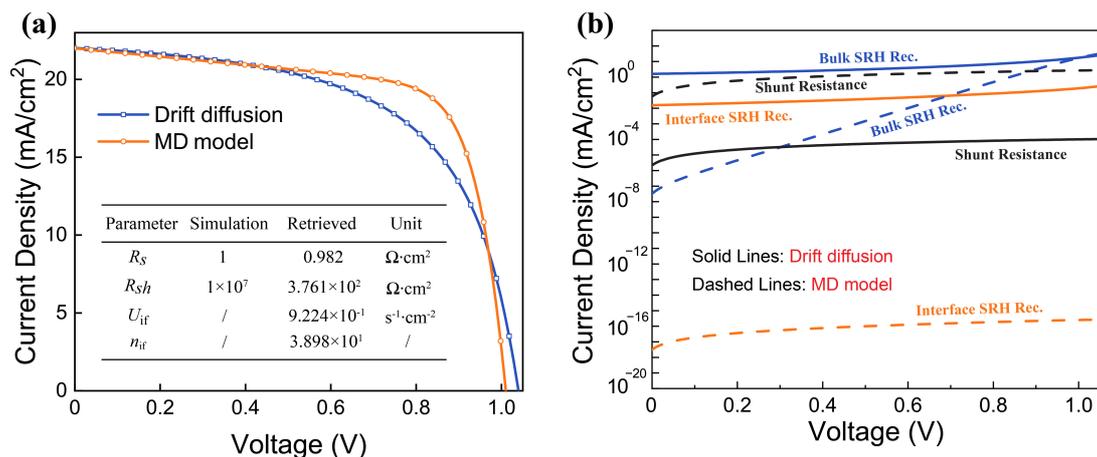

**Figure S4.** Current density-voltage relations of a device with severe bulk SRH recombination (10 ns recombination lifetime).

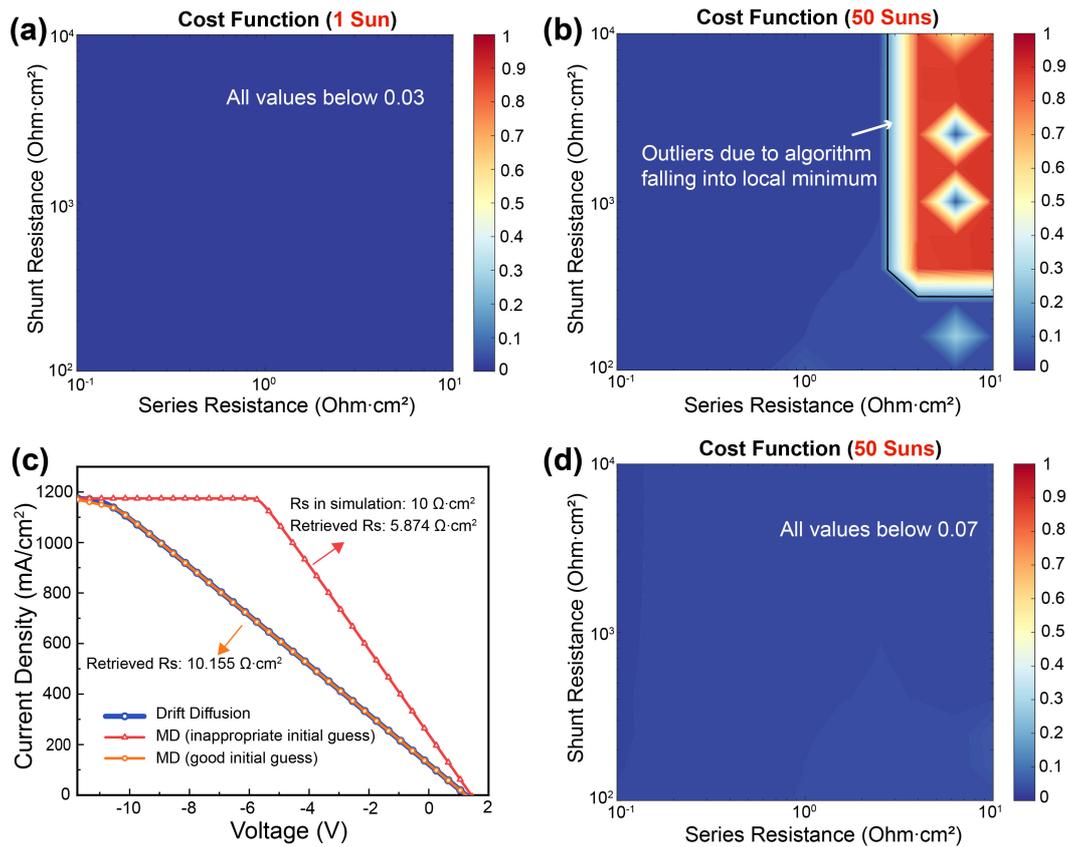

**Figure S5.** Contour plots of cost as a function of series resistance and shunt resistance. **(a-b)** is calculated by a fixed initial guess for $R_s$, $1 \ \Omega \cdot cm^2$, in the MD model. The outlier with a very high cost function is further analyzed in **(c)**, where we can see that a good initial guess for $R_s$ is critical for the accuracy of curve fitting. With this knowledge, in **(d)**, we use the retrieved values of $R_s$ in **(a)** as the initial guesses for fitting the 50 Suns curves, where uniformly low cost-function is achieved, showing that this strategy (or randomized initial guess) can help retrieve precise parameter values and validate model accuracy.

**WILEY-VCH**

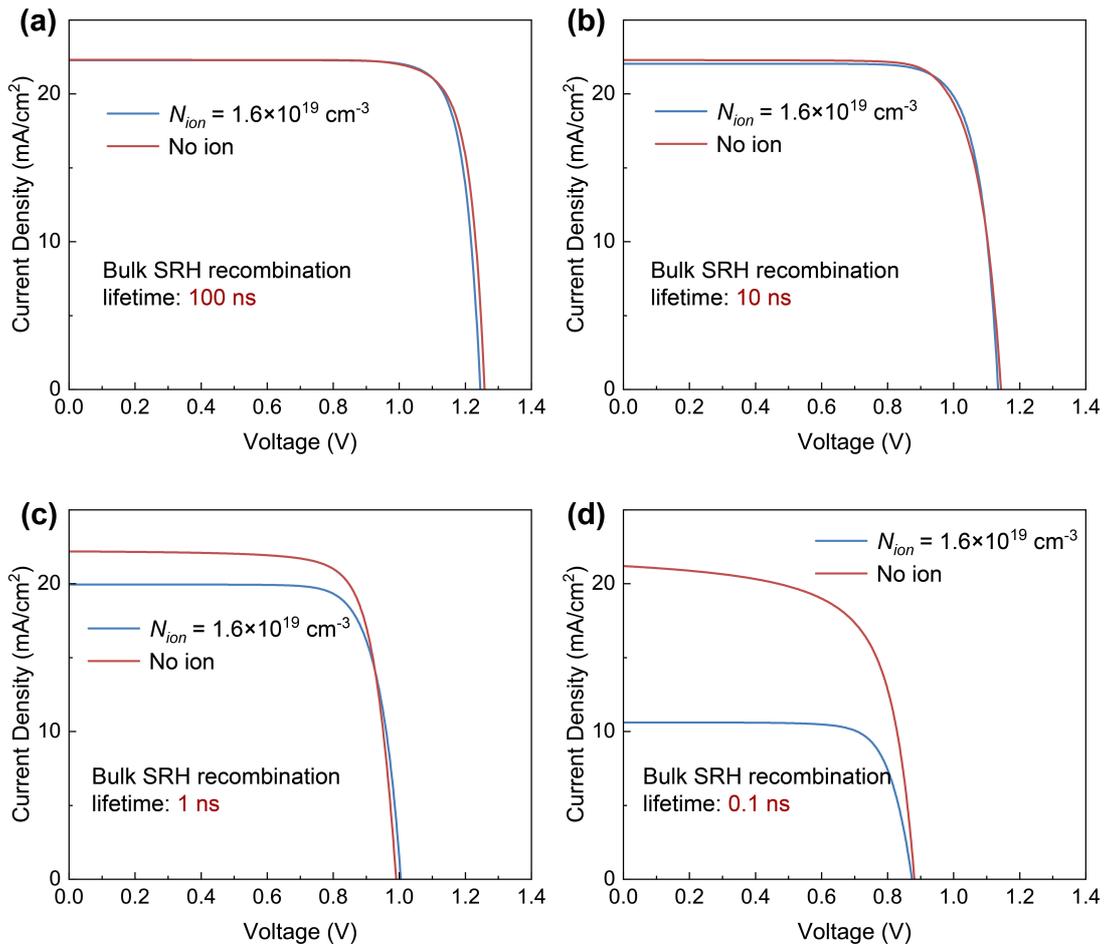

**Figure S6.** Simulated current density-voltage relations incorporating mobile ions for the device with predominant bulk SRH recombination. The bulk SRH recombination lifetimes are **(a)** 100 ns, **(b)** 10 ns, **(c)** 1 ns, and **(d)** 0.1 ns, respectively. The results are simulated with SolarDesign; identical trends can be reproduced by IonMonger.

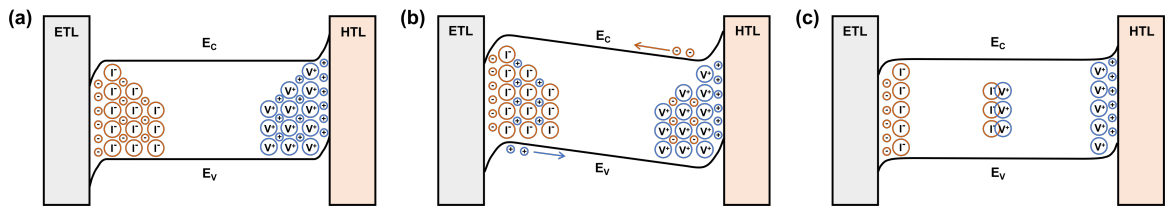

**Figure S7.** Ionic and electronic distributions and the corresponding band diagrams in perovskite photovoltaics: **(a)** in thermal equilibrium; **(b)** immediately upon illumination; **(c)** after sufficient time of illumination.

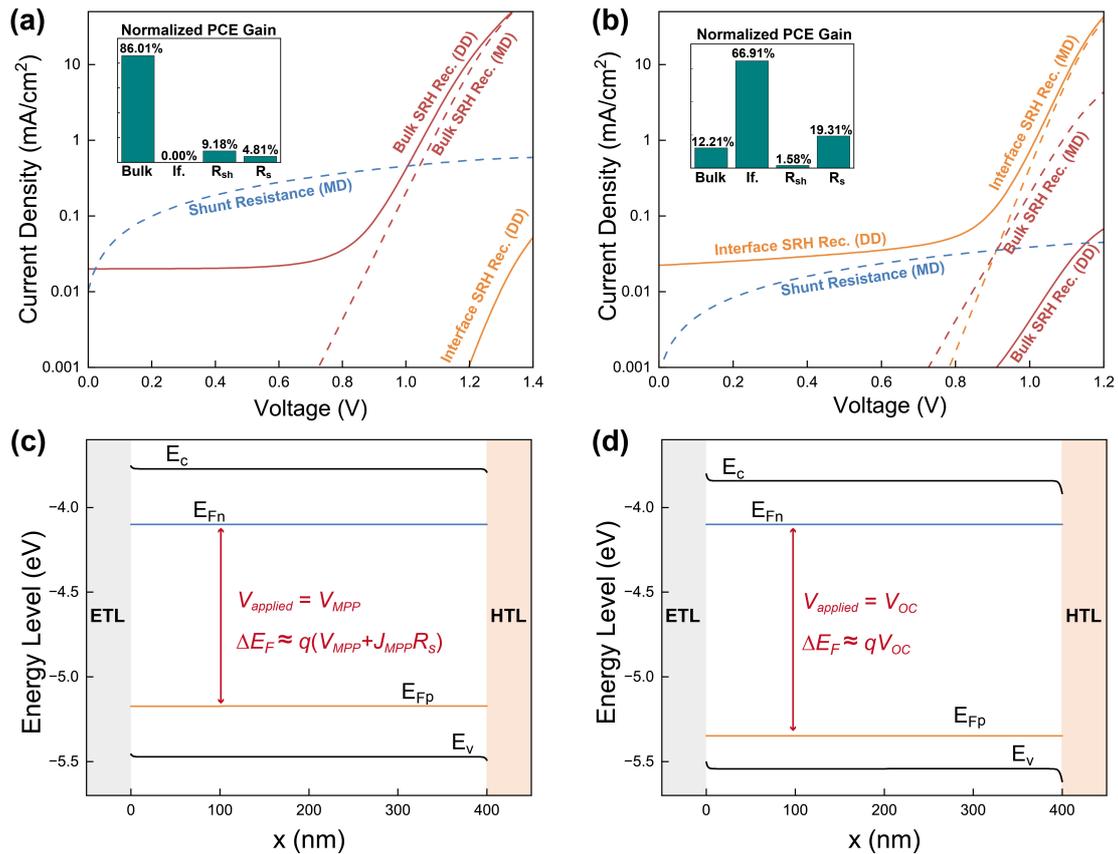

**Figure S8.** MD model fitting for steady-state $JV$ curves incorporating $10^{25}$ m$^{-3}$ ionic vacancies. **(a)** The predominant loss pathway is bulk SRH recombination. **(b)** The predominant loss pathway is interface SRH recombination. **(c)** Energy level diagram for bulk SRH recombination limited device at maximum-power-point voltage. **(d)** Energy level diagram for bulk SRH recombination limited device at open-circuit voltage.

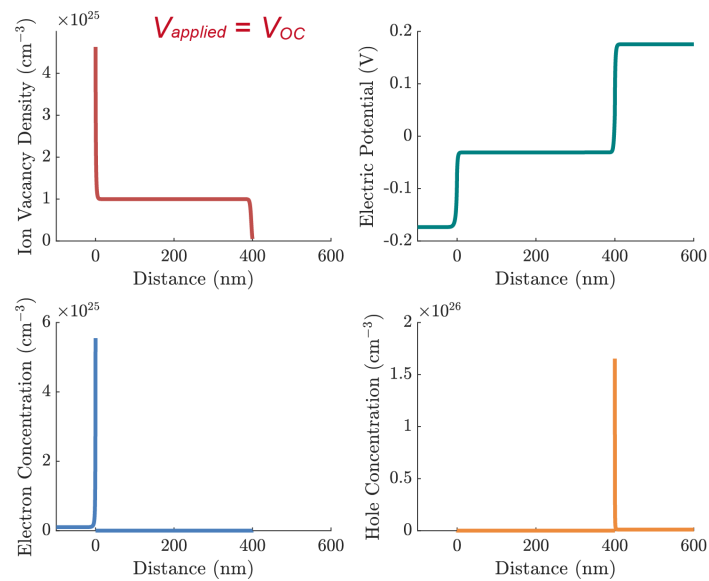

**Figure S9.** Ion vacancy density, carrier density, and electric potential distribution in the device with $10^{25}$ m$^{-3}$ ionic vacancies at open-circuit voltage.

**WILEY-VCH**

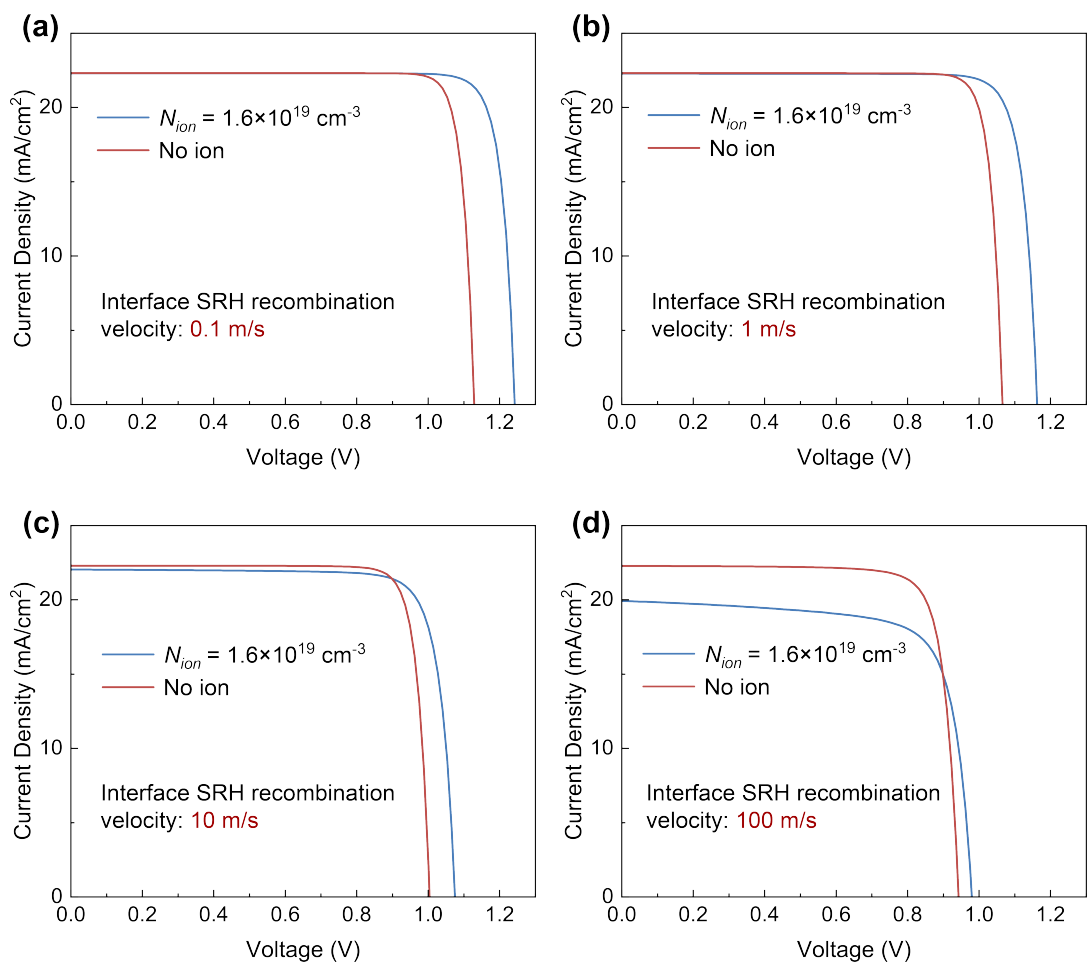

**Figure S10.** Simulated current density-voltage relations incorporating mobile ions for the device with predominant interface SRH recombination. The interface SRH recombination velocities are **(a)** 0.1 m/s, **(b)** 1 m/s, **(c)** 10 m/s, and **(d)** 100 m/s, respectively. The results are simulated with SolarDesign; identical trends can be reproduced by IonMonger.

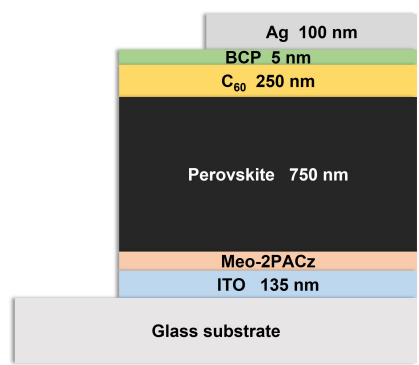

**Figure S11.** Device structure of the perovskite photovoltaics used in the experiments.

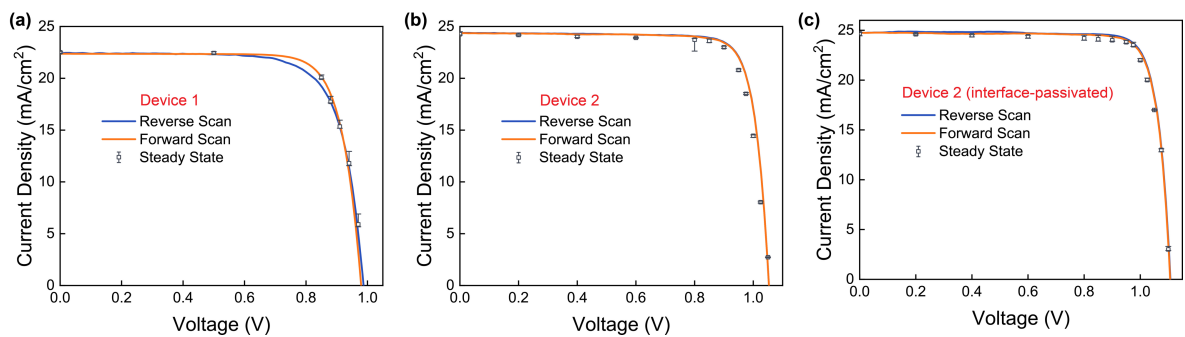

**Figure S12.** Current density-voltage relations of Device 1, Device 2, and interface-passivated Device 2. The range and average values of the recorded steady-state data points at specific voltages are also depicted in the corresponding figures.

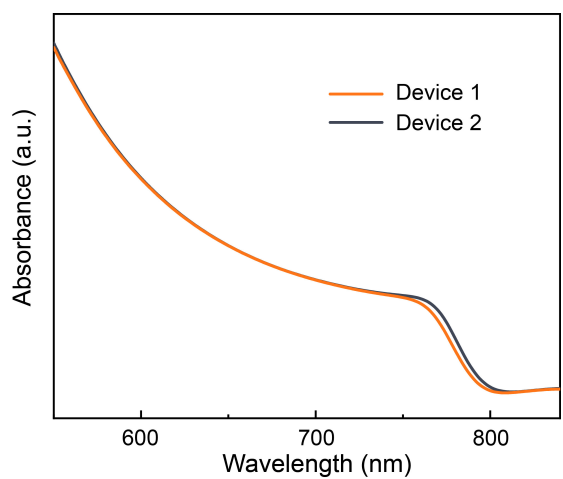

**Figure S13.** UV-vis absorbance as a function of wavelength.

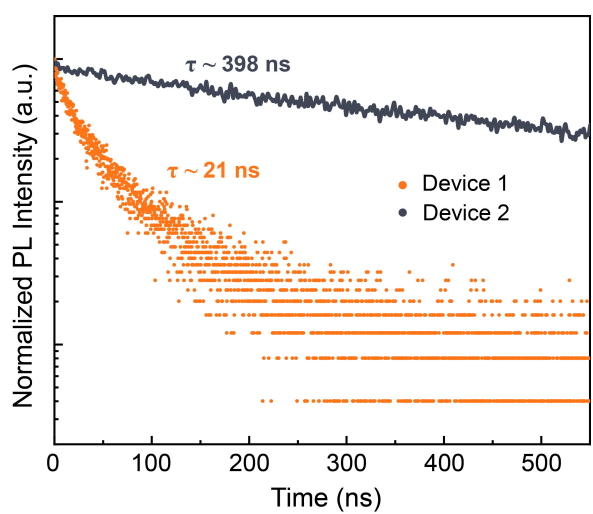

**Figure S14.** TRPL decay of the perovskite films. Device 1 and Device 2 denote that the film is made in the same condition as the corresponding device.

**WILEY-VCH**

**Table S1.** Device parameters used for drift-diffusion simulations

| Parameter | Symbol | Value | Unit |
|---|---|---|---|
| Thickness of PTAA | $d_{PTAA}$ | 10 | nm |
| Thickness of perovskite | $d_{pero}$ | 700 | nm |
| Thickness of C$_{60}$ | $d_{C60}$ | 30 | nm |
| Electron affinity of PTAA | $E_{A,PTAA}$ | 2.5 | eV |
| Bandgap of PTAA | $E_{G,PTAA}$ | 3 | eV |
| Electron affinity of perovskite | $E_{A,pero}$ | 3.9 | eV |
| Bandgap of perovskite | $E_{G,pero}$ | 1.6 | eV |
| Electron affinity of C$_{60}$ | $E_{A,C60}$ | 3.9 | eV |
| Bandgap of C$_{60}$ | $E_{G,C60}$ | 1.8 | eV |
| Work function of ITO | $W_{ITO}$ | 5.4 | eV |
| Work function of Ag | $W_{Ag}$ | 4.0 | eV |
| Relative dielectric constant of PTAA | $\epsilon_{PTAA}$ | 3.5 | / |
| Relative dielectric constant of perovskite | $\epsilon_{pero}$ | 22 | / |
| Relative dielectric constant of C$_{60}$ | $\epsilon_{C60}$ | 5 | / |
| Effective density of states in PTAA | $N_{C/V,HTL}$ | $1 \times 10^{20}$ | cm$^{-3}$ |
| Effective density of states in perovskite | $N_{C/V,pero}$ | $3.1 \times 10^{18}$ | cm$^{-3}$ |
| Effective density of states in C$_{60}$ | $N_{C/V,C60}$ | $1 \times 10^{20}$ | cm$^{-3}$ |
| Carrier mobilities in PTAA | $\mu_{PTAA}$ | $1.5 \times 10^{-4}$ | cm$^2$/(Vs) |
| Carrier mobilities in perovskite | $\mu_{pero}$ | 1 | cm$^2$/(Vs) |
| Carrier mobilities in C$_{60}$ | $\mu_{C60}$ | $1 \times 10^{-2}$ | cm$^2$/(Vs) |
| Effective doping density in PTAA | $N_A^-$ | 0 | cm$^{-3}$ |
| Effective doping density in C$_{60}$ | $N_D^+$ | 0 | cm$^{-3}$ |
| Electron lifetime for bulk SRH recombination* | $\tau_n$ | 500 | ns |
| Hole lifetime for bulk SRH recombination* | $\tau_p$ | 500 | ns |
| SRH recombination velocity at ETL/perovskite interface* | $S_{ETL/perov}$ | 2000 | cm/s |
| SRH recombination velocity at HTL/perovskite interface* | $S_{HTL/perov}$ | 200 | cm/s |
| Temperature | $T$ | 298 | K |
| Radiative recombination coefficient | $B_{rad}$ | $9.8 \times 10^{-12}$ | cm$^4$/s |
| Capture cross section | $\sigma$ | $1 \times 10^{-15}$ | cm$^2$ |

**WILEY-VCH**

| | | | |
|---|---|---|---|
| Thermal velocity | $v_{th}$ | $1 \times 10^7$ | cm/s |
| External series resistance* | $R_s$ | 1 | $\Omega \cdot cm^2$ |
| External shunt resistance* | $R_{sh}$ | $1 \times 10^7$ | $\Omega \cdot cm^2$ |

Note: The parameter values are adopted from the references[6, 7], except the thickness of the perovskite layer and the parameters marked with asterisk symbols (indicating that they can be varied during the simulations).

**Table S2.** Retrieved MD model parameters for Device 2 under laser illumination

| Irradiance (W/cm²) | Scan Direction | $R_s$ ($\Omega \cdot cm^2$) | $R_{sh}$ ($\Omega \cdot cm^2$) | $U_{if}$ ($s^{-1}cm^{-2}$) | $n_{if}$ |
|---|---|---|---|---|---|
| 0.1 | F | 0.416 | 1.343e3 | 1.719e8 | 1.930 |
| | R | 0.439 | 5.189e2 | 3.930e7 | 1.808 |
| 1.0 | F | 0.370 | 4.892e1 | 1.317e13 | 3.511 |
| | R | 0.383 | 8.607e1 | 4.805e12 | 3.239 |
| 2.0 | F | 0.380 | 3.164e1 | 2.053e13 | 3.450 |
| | R | 0.390 | 4.823e1 | 1.348e13 | 3.325 |
| 3.0 | F | 0.401 | 3.734e1 | 3.375e13 | 3.493 |
| | R | 0.392 | 3.907e1 | 2.072e13 | 3.325 |
| 4.0 | F | 0.363 | 5.189 | 1.607e13 | 3.170 |
| | R | 0.380 | 1.791e1 | 2.581e13 | 3.257 |
| 5.0 | F | 0.366 | 6.389 | 2.050e13 | 3.165 |
| | R | 0.384 | 9.737 | 1.506e13 | 3.053 |

**Table S3.** Device parameters used for drift-diffusion simulations incorporating ion migration

| Predominant Recombination | Parameter | Value | Unit |
|---|---|---|---|
| Bulk SRH recombination | Electron pseudo-lifetime for SRH | 100-0.1 | ns |
| | Hole pseudo-lifetime for SRH | 100-0.1 | ns |
| | Electron recombination velocity for SRH | $10^{-5}$ | m/s |
| | Hole recombination velocity for SRH | $10^{-5}$ | m/s |
| Interface SRH recombination | Electron pseudo-lifetime for SRH | $10^5$ | ns |
| | Hole pseudo-lifetime for SRH | $10^5$ | ns |
| | Electron recombination velocity for SRH | 0.1-100 | m/s |
| | Hole recombination velocity for SRH | 0.1-100 | m/s |

**Table S4.** Optimization strategies for different loss pathways

| Predominant Loss Pathway | Optimization Strategies |
|---|---|
| Bulk SRH recombination | - Compositional engineering<br>- Dimensionality engineering<br>- Morphology optimization<br>- Crystallization control |
| Interface SRH recombination | - Interface defect passivation<br>- Band-level alignment<br>- Morphology Control |
| Series resistance | - Metal grid<br>- Transport layer doping<br>- High mobility transparent conduction layers |
| Shunt resistance | - Reducing manufacturing defects<br>- Reducing low-voltage recombination currents |

**Table S5.** Causes to large fitting error and corresponding solutions

| Causes to large fitting error | Solutions |
|---|---|
| Low bulk SRH recombination lifetime | Referring to TRPL results |
| High interface SRH recombination velocity | Identifying contributions of shunt resistance |
| Algorithm falling into local minimum | Using random sets of initial parameters |

**Table S6.** Comparison between the MD model and classical models for photovoltaics

| Model | Strengths | Weaknesses | Applications |
|---|---|---|---|
| Drift-Diffusion Models | - Physical insight<br>- Broad applicability<br>- Steady-state and transient behavior | - Too many parameters<br>- Questionable parameter uniqueness<br>- Parameter sensitivity<br>- Numerical complexity | - Study device physics<br>- Study complex device structures<br>- Predict performance |
| Classical Diode Models | - Rapid computation<br>- Parameter extraction | - Over-simplified parameters<br>- Limited physical interpretability | - Predict performance<br>- Design PV modules and arrays |
| MD Model (this work) | - Accurate loss quantification<br>- Clear parameter physical meanings | - Domain-specific<br>- Fitting error can be large in certain cases | - Quantify losses in perovskite PVs<br>- Predict performance |

**WILEY-VCH**